\documentclass[prd,aps,showpacs,nofootinbib]{revtex4-1}%
\usepackage{graphicx}
\usepackage{amssymb}
\usepackage{amsmath}
\usepackage[dvipsnames]{xcolor}
\usepackage{float}
\usepackage{graphicx}
\usepackage{graphicx}
\usepackage{amsfonts}
\usepackage[colorlinks=true,pdfstartview=FitV,linkcolor=blue,citecolor=blue,urlcolor=blue,breaklinks=true]%
{hyperref}%
\usepackage{subfig}
\providecommand{\U}[1]{\protect\rule{.1in}{.1in}}
\begin{document}

\title{Maxwell electrodynamics modified by a \textit{CPT}-odd \\ dimension-five higher-derivative term}
\author{Manoel M. Ferreira Jr}
\email{manojr.ufma@gmail.com}
\author{Let\'{\i}cia Lisboa-Santos}
\email{let\_lisboa@hotmail.com}
\author{Roberto V. Maluf}
\email{r.v.maluf@fisica.ufc.br}
\author{Marco Schreck}
\email{marco.schreck@ufma.br}

\affiliation{Departamento de F\'{\i}sica, Universidade Federal do Maranh\~{a}o, Campus
Universit\'{a}rio do Bacanga, S\~{a}o Lu\'{\i}s, Maranh\~{a}o, 65080-805, Brazil}
\affiliation{Departamento de F\'{\i}sica, Universidade Federal do Cear\'{a}, Campus
Universit\'{a}rio do Piau\'{\i}, Caixa Postal 6030, Fortaleza, Cear\'{a}, 60455-760, Brazil}

\begin{abstract}
In this paper, we consider an electrodynamics of higher derivatives coupled to
a Lorentz-violating background tensor. Specifically, we are interested in a dimension-five
term of the \textit{CPT}-odd sector of the nonminimal Standard-Model Extension.
By a particular choice of the operator $\hat{k}_{AF}$, we obtain a higher-derivative
version of the Carroll-Field-Jackiw (CFJ) term,
$\frac{1}{2}\epsilon^{\kappa\lambda\mu\nu}A_{\lambda}D_{\kappa}\square
F_{\mu\nu}$, with a Lorentz-violating background vector $D_{\kappa}$. This
modification is subject to being investigated. We calculate the
propagator of the theory and from its poles, we analyze the dispersion relations of
the isotropic and anisotropic sectors. We verify that classical causality is valid
for all parameter choices, but that unitarity of the theory is generally not assured.
The latter is found to break down for certain configurations of the background field
and momentum. In an analog way, we also study a dimension-five anisotropic higher-derivative
CFJ term, which is written as
$\epsilon ^{\kappa\lambda\mu\nu}A_{\lambda}T_{\kappa}(T\cdot\partial)^{2}F_{\mu\nu}$
and is directly linked to the photon sector of Myers-Pospelov theory. Within the second
model, purely timelike and spacelike $T_{\kappa}$ are considered.
For the timelike choice, one mode is causal, whereas the other is noncausal. Unitarity is
conserved, in general, as long as the energy stays real --- even for the noncausal mode. For the
spacelike scenario, causality is violated when the propagation direction lies within
certain regimes. However, there are particular configurations preserving unitarity and
strong numerical indications exist that unitarity is guaranteed for all purely spacelike
configurations. The results improve our understanding of nonminimal {\em CPT}-odd
extensions of the electromagnetic sector.
\end{abstract}

\maketitle

\section{Introduction}

The minimal Standard-Model Extension (SME), which was proposed by
V.A.~Kosteleck\'{y} and S.~Samuel in 1998~\cite{Colladay,Samuel}, shares
various established properties with the Standard Model (SM) such as
power-counting renormalizability, energy-momentum conservation, and gauge
invariance. However, it does not preserve Lorentz symmetry and, beyond that,
it can violate \textit{CPT} symmetry \cite{Greenberg:2002uu}. The SME is an effective field-theory
framework obtained from the SM including additional
terms composed of observer Lorentz-invariant contractions of the physical SM fields
and fixed background tensors. Studies based on the SME have been carried out to look for
Lorentz-violating (LV) effects and to develop a precision program that
may allow us to examine the limitation of Lorentz symmetry in various
physical interactions. In this sense, a large number of investigations has
been realized in the context of the fermion sector~\cite{Fermion1,Fermion2,Fermion3},
\textit{CPT} symmetry violation \cite{CPT}, the electromagnetic
\textit{CPT}-odd sector~\cite{Adam1,Adam2,Cherenkov1}, the electromagnetic
\textit{CPT}-even sector~\cite{Cherenkov2,KM}, photon-fermion interactions
\cite{KFint,Interac,Schreck1}, and radiative corrections \cite{Radio1,Radio2,Radio3}.
Phenomenological and theoretical developments focusing on LV
contributions of mass dimensions 3 and 4 have been continuously undertaken in
the latest years. As a result, there is now a large number of tight constraints
on Lorentz violation, mainly in the photon and lepton sector \cite{KostRussell}.

To enable theoretical predictions of quantum-gravity effects at energies much smaller than
the Planck scale, we must assume that such a fundamental theory has a perturbative
expansion in terms of the dimensionless ratio $E/M_{\mathrm{QG}}$. Here, $E$ is the
energy of some experiment of interest and $M_{\mathrm{QG}}$ is the mass scale where the
full theory of quantum gravity must be considered to describe the physics as a whole.
The latter is expected to be related to the Planck mass; at least it is supposed to
lie in its vicinity: $M_{\mathrm{QG}}\simeq M_{\mathrm{Pl}}$. In configuration space,
a single power of energy corresponds to a spacetime derivative that occurs in an
appropriate field operator. Based on the existence of such an asymptotic
expansion, the SM described by the Lagrange density $\mathcal{L}_{\mathrm{sm}}$ can be
interpreted as a zeroth-order contribution of this series. The SM depends on
a range of different mass scales such as the electron mass $m_e$, the proton mass $m_p$
or the electroweak mass scale $M_{\mathrm{ew}}$. In what follows, we will choose $M_{\mathrm{ew}}$
as its characteristic mass scale.

The difference between the minimal SME and the SM is that the former
involves Lorentz-violating controlling coefficients that are either of mass dimension 1
or dimensionless. They are contracted with field operators of mass dimension 3 and 4,
respectively (incorporated in $\mathcal{L}_{\mathrm{sme}}{}^{\hspace{-0.5cm}(3,4)}$).\footnote{The standard
notation employed is to indicate the mass dimensions $d$ of the corresponding field
operators as indices in parentheses.} For dimensional reasons and to
take into account a suppression of Lorentz violation by the scale $M_{\mathrm{QG}}$,
the dimension-four coefficients could be formed via appropriate constant ratios such as
$M_{\mathrm{ew}}/M_{\mathrm{QG}}$. These contributions do not rise with energy,
which is why formally they also must be considered as of zeroth order.
The properties of the dimension-three coefficients are peculiar in this sense. Following the
previous arguments, they should be of order $-1$, i.e., they ought to
have a dependence of the form
$M_{\mathrm{QG}}/E$. Hence, they are expected to be suppressed for increasing energies, but for energies
small enough, they would be large due to the occurrence of the quantum-gravity mass scale
in the numerator. Obviously, there is no experimental evidence for Lorentz violation that
big, whereupon these terms must be suppressed in a different manner.

Taking into account contributions that grow with energy, naturally leads us
to power-counting nonrenormalizable theories. There may be Lorentz-invariant terms (contained
in $\delta\mathcal{L}_{\mathrm{LI}}^{(d)}$) that are suppressed by a mass scale $M$
not necessarily related to the Planck mass. As these terms are Lorentz-invariant, their
origin might lie in a regime different from that of Planck-scale physics. They are also known
to improve the ultraviolet behavior of theories, which was certainly one of the main
motivations for exploring such terms (see below).

In the latest years, an interplay between Lorentz violation and theories
endowed with higher derivatives has taken place. Indeed, Lorentz violation can incorporate
operators of higher mass dimensions, which may include higher-derivative terms. The
number of such contributions is infinite in contrast to the minimal LV
extensions. Nonminimal versions of the SME were developed for both the photon
\cite{Kostelec1} and the fermion sector \cite{Kostelec2,Joao}.
The Lorentz-violating contributions are supposedly suppressed by powers of the Planck scale and
are comprised by the Lagrange density $\delta\mathcal{L}_{\mathrm{sme}}{}^{\hspace{-0.5cm}(d)}$ {}
of the nonminimal SME. So we can write the asymptotic series in the form
\begin{subequations}
\label{eq:perturbative-expansion-quantum-gravity}
\begin{align}
\mathcal{L}_{\mathrm{QG}}&=\mathcal{L}_{\mathrm{sm}}+\delta\mathcal{L}_{\mathrm{sme}}^{(3)}+\delta\mathcal{L}_{\mathrm{sme}}^{(4)}+
\sum_{d\geq 5}^{\infty} \left[\delta\mathcal{L}_{\mathrm{LI}}^{(d)}+\delta\mathcal{L}_{\mathrm{sme}}^{(d)}\right]+\delta\mathcal{L}_{\mathrm{other}}\,, \displaybreak[0]\\[2ex]
\mathcal{L}_{\mathrm{sm}}&=\mathcal{L}_{\mathrm{sm}}\left(m_e,m_p,M_{\mathrm{ew}},\dots\right)\,, \displaybreak[0]\\[2ex]
\delta\mathcal{L}_{\mathrm{sme}}^{(3)}&\supset M_{\mathrm{QG}}n^X\hat{O}_X^{(3)}\,,\quad \delta\mathcal{L}_{\mathrm{sme}}^{(4)}\supset \frac{M_{\mathrm{ew}}}{M_{\mathrm{QG}}}n^X\hat{O}_X^{(4)}\,, \displaybreak[0]\\[2ex]
\delta\mathcal{L}_{\mathrm{LI}}^{(d)}&\supset\frac{1}{M^{d-4}}\hat{O}^{(d)}\,,\quad \delta\mathcal{L}_{\mathrm{sme}}^{(d)}\supset\frac{1}{M_{\mathrm{QG}}^{d-4}}n^X\hat{O}_X^{(d)}\,.
\end{align}
\end{subequations}
Here, we also stated the general structure of individual terms. A generic \textit{dimensionless} background
tensor with a set $X$ of Lorentz indices is indicated by $n^X$ where $\hat{O}_X$ is a field operator
with the same set of Lorentz indices. The mass dimension of the background has been extracted explicitly
to show the dependence on the scales $M_{\mathrm{QG}}$ and $M$. The remaining Lagrange density
$\delta\mathcal{L}_{\mathrm{other}}$ contains all additional contributions that have not been taken into
account previously. These could be nonperturbative in nature.

According to the perturbative series of Eq.~(\ref{eq:perturbative-expansion-quantum-gravity}),
the nonminimal SME is a natural extension of the higher-derivative Lorentz-invariant
contributions in the same sense as the minimal SME is an extension of the SM.
Searching for physics beyond the SM via an effective theory, there is
\emph{a priori} no reason why such nonminimal terms should be discarded. Interpreting these
terms as theories that are valid within a certain energy range only, power-counting
nonrenormalizability is not considered to be a problem.

Due to the arguments explained above, individual terms have an energy dependence of the form
$(E/M_{\mathrm{QG}})^{d-4}$ and $(E/M)^{d-4}$, respectively. Nonminimal contributions grow with energy,
i.e., after a certain point they dominate the minimal ones. Therefore, they might be essential in
experiments involving particles of high energies, e.g., cosmic rays. From the theoretical perspective,
fundamental properties such as causality, stability, and unitarity could be investigated in the high-energy
regime where the higher-derivative field operators become dominant. In what follows, we will state several examples for nonrenormalizable
extensions of electrodynamics that are Lorentz-invariant.

The first example of an extended electrodynamics including higher
derivatives was proposed in 1942 by Podolsky~\cite{Podolsky1}. He initially studied the Lorentz- and
gauge-invariant dimension-6 term, $\theta^{2}\partial_{\alpha}F^{\alpha
\beta}\partial_{\lambda}F_{\phantom{\lambda}\beta}^{\lambda}$, with
the Podolsky parameter $\theta$ of mass dimension $-1$. This theory exhibits two
dispersion relations, the usual one of Maxwell theory and a massive mode that
renders the self-energy of a pointlike charge finite. However, at the quantum level,
the massive mode produces ghosts \cite{Accioly1}. The gauge-fixing condition for this
extension can be adapted to be compatible with the 2 degrees of freedom of the photon
and the 3 additional ones connected to the massive mode \cite{GalvaoP}.
Further developments \cite{Podolsky3,Podolsky5} in Podolsky's theory deserve to be mentioned
including investigations of a dimension-six quantum electrodynamics in (1+1)
spacetime dimensions \cite{Manavella}.

Another relevant extension of Maxwell theory with higher derivatives is Lee-Wick
electrodynamics, described by the dimension-6 term $F_{\mu\nu
}\partial_{\alpha}\partial^{\alpha}F^{\mu\nu}$ \cite{LeeWick}. This theory
also implies a finite self-energy for a pointlike charge in $(1+3)$ spacetime
dimensions. It provides a bilinear contribution to the Maxwell Lagrangian that
is similar to the Podolsky term but with opposite sign. This term causes
energy instabilities at the classical level and negative-norm states in the Hilbert
space at the quantum level. Lee and Wick introduced a mechanism to
preserve unitarity by removing all states with negative norm from the Hilbert
space. Ten years ago, this theory received attention again with the
proposal of the Lee-Wick Standard Model \cite{LW}, which is based on a
non-Abelian gauge structure free of quadratic divergences. The Lee-Wick Standard Model
has had a big impact and is applied in both theory and phenomenology~\cite{LW2}.
In this context, general investigations of LV extensions are found in \cite{Turcati}
as well as applications to the self-energy and interaction of pointlike and spatially
extended sources \cite{Barone2,Accioly2,Barone1,Barone3}. General
considerations of the pole structure and perturbative unitarity of these classes of
theories can be found in \cite{Anselmi}.

As we explained above, theories of higher derivatives and higher-derivative electrodynamics, in particular,
can be endowed with Lorentz-violating contributions. It is important to mention that special choices
of nonminimal operators were proposed and investigated \cite{Myers1,Marat,Reyes:2010pv,Reyes:2013nca}
where some works such as \cite{Passos} even involve Ho\v{r}ava-Lifshitz gravity.
Nonminimal theories containing higher-dimensional couplings can also be constructed
without introducing higher derivatives (beyond those contained inside the field strength tensor). Such couplings were considered
and constrained initially in \cite{NModd1,NM3} and have been proposed recently
in broader scenarios \cite{NMkost, NMkost2}, also including photon-photon scattering
\cite{Gomes}, electroweak interactions \cite{VictorEW}, nuclear chiral interactions
\cite{VictorCar}, scalar electrodynamics \cite{Josberg}, and scattering
processes of electrons and positrons \cite{Khanna}.

Another motivation to studying higher-derivative Lorentz-violating terms are noncommutative
spacetime theories. These are based on commutation relations for spacetime coordinates
of the form $[x^{\mu},x^{\nu}]=\theta^{\mu\nu}$. The object $\theta^{\mu\nu}$ is fixed
with respect to particle Lorentz transformations and is interpreted as an observer
tensor giving rise to preferred spacetime directions. Due to dimensional reasons,
$[\theta^{\mu\nu}]=-2$, whereupon it is inevitable that this object must be linked
to terms of the nonminimal SME. This property was shown explicitly in \cite{Carroll:2001ws}
by applying the Seiberg-Witten map to translate a noncommutative field theory into
a commutative field theory of modified photons.

Lately, modified higher-derivative LV terms have been generated in
other ways via quantum corrections to the photon effective action in
a scenario with a nonminimal coupling between fermions and photons
\cite{PetrovPLB2016,Barone4} as well as in supersymmetric theories
\cite{Bonetti1,Bonetti2}. Recently, dimension-five terms of Myers-Pospelov
type have been investigated in the context of black-body radiation \cite{Anacleto}
and emission of electromagnetic and gravitational waves \cite{Anacleto2}. Higher-derivative
applications to gamma rays have also been taken into consideration \cite{Gamma}.
Not so long ago, we addressed a \textit{CPT}-even,
dimension-6, higher-derivative electrodynamics, composed of an anisotropic
Podolsky term, $\partial_{\sigma}F^{\sigma}{}^{\beta}\partial_{\lambda}F^{\lambda}{}%
^{\alpha}D_{\beta\alpha}$, and an anisotropic Lee-Wick term,
$F_{\mu\nu}\partial_{\alpha}\partial_{\beta}F^{\mu}{}^{\nu}D^{\alpha\beta}$,
respectively, with $D^{\alpha\beta}$ representing a LV rank-2 background field. Both models
can be mapped onto dimension-6 terms of the electromagnetic sector within the SME.
We obtained the associated gauge propagators and examined the dispersion relations
for several background tensor configurations. We found that both models exhibit
both causal and noncausal as well as both unitary and nonunitary modes
\cite{Leticia1}.

To improve our understanding of higher-derivative Lorentz-violating extensions
of the photon sector, it is now reasonable to pursue similar investigations of a
{\em CPT}-odd theory. Therefore, in the present work, we study Maxwell electrodynamics modified by
a \textit{CPT}-odd, dimension-five nonminimal SME term. In Sec.~\ref{sec:modification-simple-model}, we consider
a configuration composed of a fixed background field $D_{\kappa}$ where the additional
four-derivatives are contracted with the metric tensor, i.e., $\epsilon^{\kappa\lambda\mu\nu}A_{\lambda}D_{\kappa
}\square F_{\mu\nu}$. The gauge propagator is derived and the dispersion relations
are obtained from its pole structure. Causality and unitarity of the modes is
analyzed subsequently. In Sec.~\ref{sec:modification-sophisticated-model}, we analyze
an alternative configuration composed of a
fixed background field $T_{\kappa}$ partially contracted with additional four-derivatives,
that is, $\epsilon^{\kappa\lambda\mu\nu}A_{\lambda}T_{\kappa}(T\cdot\partial)^{2}F_{\mu\nu}$.
Similar investigations are performed for this more sophisticated modification that
is linked to the photon sector of Myers-Pospelov theory. Finally,
we conclude on our findings in Sec.~\ref{sec:conclusions}. Studies that are not directly
connected to the modifications proposed are relegated to Appx.~\ref{sec:comparison-propagator-mcfj-theory},
\ref{sec:mapping-nonminimal-sme}.
Natural units will be used with $\hbar=c=1$, unless otherwise stated.

\section{Maxwell electrodynamics modified by a CPT-odd dimension-five higher-derivative
term: a simple model}
\label{sec:modification-simple-model}

The Lagrange density for the nonminimal SME photon sector \cite{Kostelec1} is
written in a way similar to that of the minimal sector:
\begin{equation}
\mathcal{L}_{\gamma}=-\frac{1}{4}F_{\mu\nu}F^{\mu\nu}+\frac{1}{2}%
\epsilon^{\kappa\lambda\mu\nu}A_{\lambda}(\hat{k}_{AF})_{\kappa}F_{\mu\nu
}-\frac{1}{2}F_{\kappa\lambda}(\hat{k}_{F})^{\kappa\lambda\mu\nu}F_{\mu\nu}\,,
\end{equation}
with the \textit{U}(1) gauge field $A_{\mu}$ and the associated field
strength tensor $F_{\mu\nu}=\partial_{\mu}A_{\nu}-\partial_{\nu}A_{\mu}$.
The four-dimensional Levi-Civita symbol is denoted by $\epsilon^{\mu\nu\rho\sigma}$ where we
use the convention $\epsilon^{0123}=1$. All
fields are defined in Minkowski spacetime with metric tensor $(\eta_{\mu\nu
})=\mathrm{diag}(1,-1,-1,-1)$. The operators $(\hat{k}_{AF})_{\kappa}$ and $(\hat{k}_{F})^{\kappa
\lambda\mu\nu}$ now represent the nonminimal versions of the corresponding
minimal coefficients $(k_{AF})_{\kappa}$ and $(k_{F})^{\kappa\lambda\mu\nu}$.
They involve terms of higher derivatives and are given by the following
infinite \textit{CPT}-odd and \textit{CPT}-even operator series:
\begin{subequations}
\begin{align}
(\hat{k}_{AF})_{\kappa}&=\sum\limits_{d \text{ odd}}
(k_{AF}^{(d)})_{\kappa}^{\phantom{\kappa}\alpha_{1}\ldots\alpha_{(d-3)}%
}\partial_{\alpha_{1}}\ldots\partial_{\alpha_{(d-3)}}\,, \displaybreak[0]\label{KAFnm}%
\\[0.03in]
(\hat{k}_{F})^{\kappa\lambda\mu\nu}&=\sum\limits_{d \text{ even}}
(k_{F}^{(d)})^{\kappa\lambda\mu\nu\alpha_{1}\ldots\alpha_{(d-4)}}%
\partial_{\alpha_{1}}\ldots\partial_{\alpha_{(d-4)}}\,. \label{KFnm}%
\end{align}
\end{subequations}
Each controlling coefficient is labeled by the mass dimension $d$ of the
associated field operator and the Lorentz
indices $\alpha_i$ are associated with the spacetime derivatives.
In this work, we are interested in investigating the \textit{CPT}-odd,
dimension-five extension of the electromagnetic sector that is
specifically represented by the Carroll-Field-Jackiw-like (CFJ-like) term of the Lagrange density:
\begin{equation}
\frac{1}{2}\epsilon^{\kappa\lambda\mu\nu}A_{\lambda}(\hat{k}_{AF})_{\kappa
}F_{\mu\nu}\,.
\end{equation}
The background field is a third-rank observer Lorentz tensor whose
general structure is
\begin{equation}
(\hat{k}_{AF})_{\kappa}=(k_{AF}^{(5)})_{\kappa}^{\phantom{\kappa}\alpha
_{1}\alpha_{2}}\partial_{\alpha_{1}}\partial_{\alpha_{2}}\,.
\end{equation}
As a first investigation, we consider the special case
\begin{equation}
(\hat{k}_{AF})_{\kappa}=\tilde{D}_{\kappa}X^{\alpha_{1}\alpha_{2}}%
\partial_{\alpha_{1}}\partial_{\alpha_{2}}\,,\label{eq:background-vector}%
\end{equation}
with a Lorentz-violating four-vector $\tilde{D}_{\kappa}$. The tensor
structure is chosen such that the vector properties of $(\hat{k}_{AF})_{\kappa}$ are
described by the preferred spacetime direction $\tilde{D}_{\kappa}$, whereas
the nonminimal sector is separately parameterized by the symmetric
tensor $X^{\mu\nu}$. Thus, the Maxwell-Carroll-Field-Jackiw-like (MCFJ-like) Lagrange density to be studied is
\begin{equation}
\mathcal{L}=-\frac{1}{4}F_{\mu\nu}F^{\mu\nu}+\frac{1}{2}\epsilon
^{\kappa\lambda\mu\nu}A_{\lambda}\tilde{D}_{\kappa}X^{\alpha_{1}\alpha_{2}%
}\partial_{\alpha_{1}}\partial_{\alpha_{2}}F_{\mu\nu}+\frac{1}{2\xi}%
(\partial_{\mu}A^{\mu})^{2}\,,\label{L}%
\end{equation}
with gauge fixing parameter $\xi$, i.e., the final term is included to fix the
gauge. We can consider an observer frame where $X^{\mu\nu}$ is
diagonal. In this context, the simplest case is that of a tensor $X^{\mu\nu}$
with equal spacelike coefficients. Then $X^{\mu\nu}$ is composed of a
traceless part $\overline{X}^{\mu\nu}$ and a part with nonvanishing trace that must
be proportional to the Minkowski metric tensor:
\begin{subequations}
\begin{align}
X^{\mu\nu}&=\alpha_{\mathrm{tr}}\overline{X}^{\mu\nu}+\alpha\eta^{\mu\nu
}\,, \displaybreak[0]\\[2ex]
\overline{X}^{\mu\nu}&=\mathrm{diag}(1,1/3,1/3,1/3)^{\mu\nu}=\frac{1}{3}\left[4\xi^{\mu}\xi^{\nu}-\eta^{\mu\nu}\right]\,,
\end{align}
with the preferred purely timelike direction $(\xi^{\mu})=(1,0,0,0)$ and
parameters $\alpha_{\mathrm{tr}}$, $\alpha$ suitably chosen.\footnote{Here
we simply adopt the notation used in the \textit{CPT}-even
photon sector where the single coefficient $\kappa_{\mathrm{tr}}$ is linked to
a symmetric, traceless matrix, as well.} As the traceless part involves two
preferred directions $\xi^{\mu}$ and $\tilde{D}^{\mu}$, its investigation is probably
more complicated than that of the contribution proportional to the trace.
Therefore, we leave the analysis of the traceless part for the future and
choose $\alpha_{\mathrm{tr}}=0$ so that $\tilde{D}_{\kappa}X^{\alpha_{1}\alpha_{2}}\partial
_{\alpha_{1}}\partial_{\alpha_{2}}=D_{\kappa}\square$, with the d'Alembertian $\square
\equiv\partial_{\mu}\partial^{\mu}$ and the redefined background vector
$D_{\kappa}\equiv\alpha\tilde{D}_{\kappa}$. Thus, the LV background is
\begin{equation}
(\hat{k}_{AF})_{\kappa}=\tilde{D}_{\kappa}\square\,,\label{eq:background-vector2A}%
\end{equation}
and the new Lagrange density has the form
\end{subequations}
\begin{equation}
\mathcal{L}=-\frac{1}{4}F_{\mu\nu}F^{\mu\nu}+\frac{1}{2}\epsilon
^{\kappa\lambda\mu\nu}D_{\kappa}A_{\lambda}\square F_{\mu\nu}+\frac{1}{2\xi
}(\partial_{\mu}A^{\mu})^{2}\,.\label{Lg}%
\end{equation}
By performing suitable partial integrations and neglecting boundary terms, the
latter can be written as
\begin{subequations}
\begin{equation}
\mathcal{L}=\frac{1}{2}A^{\mu}O_{\mu\nu}A^{\nu}\,,
\end{equation}
with the differential operator
\begin{equation}
O_{\mu\nu}=\square\left(  \Theta_{\mu\nu}-2L_{\mu\nu}-\frac{1}{\xi}\Omega
_{\mu\nu}\right)  \,,
\label{eq:differential-operator-o}
\end{equation}
\end{subequations}
sandwiched in between two vector fields. Here we introduced the symmetric
transversal and longitudinal projectors, $\Theta_{\mu\nu}$ and $\Omega_{\mu
\nu}$, respectively:%
\begin{equation}
\Theta_{\mu\nu}\equiv\eta_{\mu\nu}-\Omega_{\mu\nu}\,,\quad\Omega_{\mu\nu}%
\equiv\frac{\partial_{\mu}\partial_{\nu}}{\square}\,,\label{Proj1}%
\end{equation}
while the Lorentz-violating part is described by the antisymmetric and
dimensionless operator,
\begin{equation}
L_{\mu\nu}\equiv\epsilon_{\mu\nu\kappa\lambda}D^{\kappa}\partial^{\lambda}\,.
\end{equation}
Now we intend to evaluate the propagator of the theory, i.e., we should find
the Green's function $\Delta_{\alpha\beta}$, which is the inverse of the
differential operator $O_{\mu\nu}$, from the condition
\begin{equation}
O_{\mu\sigma}\Delta_{\phantom{\sigma}\nu}^{\sigma}=\eta_{\mu\nu}%
\,.\label{eq:condition-inverting-operator}%
\end{equation}
For the inverse to be found, we use a suitable basis of
tensor operators, a procedure that turned out to be successful in several theoretical
scenarios \cite{Propagator,Propagator2}. Thus, we propose
the following \textit{Ansatz}:
\begin{equation}
\Delta_{\phantom{\sigma}\nu}^{\sigma}=a\Theta_{\phantom{\sigma}\nu}^{\sigma
}+bL_{\phantom{\sigma}\nu}^{\sigma}+c\Omega_{\phantom{\sigma}\nu}^{\sigma
}+dD^{\sigma}D_{\nu}+e(D^{\sigma}\partial_{\nu}+D_{\nu}\partial^{\sigma})\,,
\end{equation}
where the parameters $a\dots e$ are expected to be scalar operators.
The algebra of the individual tensor operators is displayed in
Table~\ref{tab:closed-algebra}. We also use the definition
\begin{equation}
\rho\equiv D^{\mu}\partial_{\mu}\,,
\end{equation}
and, for brevity, define the tensor
\begin{equation}
\Gamma_{\mu\sigma}\equiv L_{\mu\nu}L_{\phantom{\nu}\sigma}^{\nu}=(D_{\mu
}\partial_{\sigma}+D_{\sigma}\partial_{\mu})\rho-D_{\mu}D_{\sigma}%
\square+(D^{2}\square-\rho^{2})\Theta_{\mu\sigma}-\rho^{2}\Omega_{\mu\sigma
}\,,\label{Gamma}
\end{equation}
which originates from the contraction of two Levi-Civita symbols.
\begin{table}[ptb]
\centering%
\begin{tabular}[c]{cccccccc}
\toprule & $\Theta_{\phantom{\sigma}\nu}^{\sigma}$ & $L_{\phantom{\sigma}\nu
}^{\sigma}$ & $\Omega_{\phantom{\sigma}\nu}^{\sigma}$ & $L_{\nu}%
^{\phantom{\nu}\sigma}$ & $D^{\sigma}D_{\nu}$ & $D^{\sigma}\partial_{\nu}$ &
$D_{\nu}\partial^{\sigma}$ \\
\colrule
$\Theta_{\mu\sigma}$ & $\Theta_{\mu\nu}$ & $L_{\mu\nu}$ & 0 & $L_{\nu\mu}$ &
$D_{\mu}D_{\nu}-\rho D_{\nu}\partial_{\mu}/\square$ & $D_{\mu}\partial_{\nu
}-\Omega_{\mu\nu}\rho$ & 0 \\
\colrule
$L_{\sigma\mu}$ & $L_{\nu\mu}$ & $-\Gamma_{\nu\mu}$ & 0 & $\Gamma_{\nu\mu}$ &
0 & 0 & 0 \\
\colrule
$L_{\mu\sigma}$ & $L_{\mu\nu}$ & $\Gamma_{\nu\mu}$ & 0 & $-\Gamma_{\nu\mu}$ & 0 & 0 &
0 \\
\colrule
$\Omega_{\mu\sigma}$ & 0 & 0 & $\Omega_{\mu\nu}$ & 0 & $\rho D_{\nu}\partial_{\mu
}/\square$ & $\Omega_{\mu\nu}\rho$ & $D_{\nu}\partial_{\mu}$ \\
\colrule
$D_{\mu}D_{\sigma}$ & $D_{\mu}D_{\nu}-\rho D_{\mu}\partial_{\nu}/\square$
& 0 & $\rho D_{\mu}\partial_{\nu}/\square$ & 0 & $D^{2}D_{\mu}D_{\nu}$ &
$D^{2}D_{\mu}\partial_{\nu}$ & $\rho D_{\mu}D_{\nu}$ \\
\colrule
$D_{\mu}\partial_{\sigma}$ & 0 & 0 & $D_{\mu}\partial_{\nu}$ & 0 & $\rho
D_{\mu}D_{\nu}$ & $\rho D_{\mu}\partial_{\nu}$ & $\square D_{\mu}D_{\nu}$ \\
\colrule
$D_{\sigma}\partial_{\mu}$ & $D_{\nu}\partial_{\mu}-\rho\Omega_{\mu\nu}$
& 0 & $\rho\Omega_{\mu\nu}$ & 0 & $D^{2}D_{\nu}\partial_{\mu}$ & $\square
D^{2}\Omega_{\mu\nu}$ & $\rho D_{\nu}\partial_{\mu}$ \\
\botrule
\end{tabular}
\caption{Closed algebra of tensor operators.}%
\label{tab:closed-algebra}%
\end{table}
Starting from the tensor equation (\ref{eq:condition-inverting-operator}),
after performing some simplifications, we have
\begin{align}
\Theta_{\mu\sigma}+\Omega_{\mu\sigma} &  \overset{!}{=}\square\left\{  \left[
a-4(D^{2}\square-\rho^{2})b\right]  \Theta_{\mu\sigma}+2(b-a)L_{\mu\sigma
}-\left[  -4\rho^{2}b+\frac{c}{\xi}+\left(  1+\frac{1}{\xi}\right)  \rho
e\right]  \Omega_{\mu\sigma}\right.  \nonumber\\
&  \phantom{{}={}\square\Big\{}\left.  +\,(d+4\square b)D_{\mu}D_{\sigma}+(e-4\rho b)D_{\mu
}\partial_{\sigma}-\left[  4\rho b+\left(  1+\frac{1}{\xi}\right)
\frac{\rho}{\square}d+\frac{1}{\xi}e\right]  D_{\sigma}\partial_{\mu}\right\}
\,.\label{S1}%
\end{align}
By comparing both sides of Eq.~(\ref{S1}) to each other, the following
differential operators are obtained:
\begin{subequations}
\begin{align}
a &  =\frac{b}{2}=\frac{1}{\boxtimes}\,,\displaybreak[0]\\[2ex]
c &  =-\left(  \frac{\xi}{\square}+\frac{4\rho^{2}}{\boxtimes}\right)
\,,\displaybreak[0]\\[2ex]
d &  =-\frac{4\square}{\boxtimes}\,,\quad e=\frac{4\rho}{\boxtimes
}\,,\displaybreak[0]\\[2ex]
\boxtimes &  =\square\left[  1+4(\rho^{2}-D^{2}\square)\right]  \,.
\end{align}
\end{subequations}
Thus, the inverse of $\mathcal{O}^{\mu\nu}$ is
\begin{equation}
\Delta_{\sigma\nu}=\frac{1}{\boxtimes}\left\{\eta_{\sigma\nu}+2L_{\sigma\nu
}-\left(  \frac{\boxtimes\xi}{\square}+1+4\rho^{2}\right)  \Omega
_{\sigma\nu}-\,4\square D_{\sigma}D_{\nu}+4\rho[D_{\sigma}\partial_{\nu
}+D_{\nu}\partial_{\sigma}]\,\right\}\,.
\end{equation}
The form of the propagator in momentum space follows from the latter result by
carrying out the substitution $\partial_{\mu}=-\mathrm{i}p_{\mu}$ with the
four-momentum $p_{\mu}$, so that:
\begin{subequations}
\label{eq:propagator-momentum-space}
\begin{align}
\Delta_{\mu\sigma}(p) &  =\frac{-\mathrm{i}}{p^{2}(1+4\Upsilon(p))}%
\bigg\{\eta_{\mu\sigma}-2\mathrm{i}\epsilon_{\mu\sigma\kappa\lambda}D^{\kappa
}p^{\lambda}-\left[1-4(D\cdot p)^{2}+\xi\left(  1+4\Upsilon(p)\right)
\right]  \frac{p_{\mu}p_{\sigma}}{p^{2}}\nonumber\\
&  \phantom{{}={}}\hspace{2.5cm}+4p^{2}D_{\mu}D_{\sigma}-4(D\cdot p)\left[
D_{\mu}p_{\sigma}+D_{\sigma}p_{\mu}\right]  \bigg\}\,,\label{propagator}%
\end{align}
with
\begin{equation}
\label{eq:propagator-quantity}
\Upsilon(p)=D^{2}p^{2}-(D\cdot p)^{2}\,.
\end{equation}
\end{subequations}
A prefactor of $\mathrm{i}$ has been added to match the conventions for the photon
propagator for zero Lorentz violation in~\cite{Peskin:1995}. Several remarks are in order.
First, in App.~\ref{sec:comparison-propagator-mcfj-theory} we present the propagator
of MCFJ theory in Lorenz gauge, showing that the latter and Eq.~(\ref{eq:propagator-momentum-space}) are linked to each other
by a simple replacement. Second, the propagator (\ref{eq:propagator-momentum-space}) is transverse, except for a piece that depends on the gauge-fixing parameter:
\begin{equation}
\label{eq:transversality-propagator}
\Delta_{\mu\sigma}(p)p^{\sigma}=\mathrm{i}\xi\frac{p_{\mu}}{p^2}\,.
\end{equation}
Third, note that contributions of the form $\Upsilon(p)$ (with the four-momentum
replaced by the four-velocity $u^{\mu}$) appear in the context of certain
classical Lagrangians associated with the SME, especially in those that arise
for the $b$ coefficients (connected to Finsler $b$ space) \cite{Finsler}.
In Euclidean space, this quantity corresponds to
the Gramian of the two vectors that appear in the expression. As there is a
connection between the \textit{CPT}-odd electromagnetic sector and the $b$
coefficients of the fermionic sector, such quantities are expected to appear.
The structure of $\Upsilon(p)$ was also observed in the graviton propagator
evaluated in the context of the linearized Einstein-Hilbert gravity (without
torsion) modified by a spontaneous violation of Lorentz symmetry. The latter
is induced by the bumblebee field \cite{MalufGravity,MalufGravity2} using
generalized versions of the Barnes-Rivers spin operator basis \cite{Barnes,Rivers,Sezgin}.
Finally, note that $\Upsilon(p)$ is a dimensionless function, as the mass
dimension of $p^{\mu}$ cancels the inverse mass dimension of $D^{\mu}$.

Fourth, it is interesting to recall that the propagator (\ref{propagator})
has an antisymmetric term, proportional to the projector $L_{\mu\sigma}$ while
the other pieces are symmetric with respect to the interchanges $\mu\rightarrow\sigma$,
$\sigma\rightarrow\mu$. The Feynman propagator\footnote{Note that we did not
determine the Feynman propagator here, as the latter requires the definition
of a suitable pole structure. The Feynman propagator is mentioned to emphasize
the symmetries that must be the same as those of the Green's function.} is defined by the vacuum expectation
value of the time-ordered product of field operators evaluated at distinct
spacetime points,
\begin{equation}
\mathrm{i}(D_{F})_{\alpha\beta}(x-y)\equiv\langle0|T(A_{\alpha}(x)A_{\beta
}(y))|0\rangle\,.
\end{equation}
The Fourier transform of the latter is symmetric with respect to the combination
of interchanging its indices and $p^{\mu}\mapsto -p^{\mu}$. The antisymmetric piece of the propagator
(\ref{propagator}) appears in \textit{CPT}-odd electrodynamics, e.g., in the MCFJ model
or Chern-Simons theory in (1+2) dimensions. But this piece does not mean that the propagator
loses its symmetry. Indeed, the propagator continues being symmetric with regards to the
two simultaneous operations mentioned before.

\subsection{Dispersion relations}

The poles of the propagator provide two dispersion equations for this model,
\begin{subequations}
\label{DR}%
\begin{align}
p^{2}&=0\,, \label{DR1}\\
1+4\left[  D^{2}p^{2}-(D\cdot p)^{2}\right]&=0\,, \label{DR2}%
\end{align}
as usual in theories with higher-dimensional operators. The
first one corresponds to the typical Maxwell pole, which also appears
in the Podolsky and Lee-Wick models as well as in the corresponding anisotropic
LV versions \cite{Leticia1}. The second equation contains information on the
higher-derivative dimension-five term. It is reasonable to compare it to the
dispersion equation obtained for the dimension-three
MCFJ theory (see Eq.~(25) in the first paper of \cite{Adam1}),
given in terms of the CFJ background vector $(k_{AF})^{\mu}$:
\end{subequations}
\begin{equation}
p^{4}+p^{2}k_{AF}^{2}-(k_{AF}\cdot p)^{2}=0\,. \label{DRodd}%
\end{equation}
The latter is a quartic-order dispersion equation, whereas
Eq.~(\ref{DR2}) is simpler (only of second order). Such a
comparison reveals that the present dimension-five MCFJ-like theory is totally
distinct from the dimension-three MCFJ model. The less involved dispersion
equation can be ascribed to the simple structure of the background
tensor that we have chosen in Eq.~(\ref{eq:background-vector}).

We will analyze the dispersion relations (DRs) for several configurations of
the LV background where it makes sense to distinguish between purely timelike
and spacelike preferred directions. If a DR does not approach
the limit of standard electromagnetic waves for zero Lorentz violation, it will
be called ``exotic.'' The latter are not necessarily unphysical, but they
decouple in the low-energy regime. Dispersion laws whose group/front
velocities are singular or exceed the value 1 will be referred to as ``spurious.''
For a purely timelike background, $D_{\gamma}=(D_{0},0)_{\gamma}$,
we have%
\begin{equation}
\mathbf{p}^{2}=\frac{D_{0}^{2}}{4}\,,
\end{equation}
which does not correspond to a propagating mode. It is a nonphysical DR, as it
does not represent a relation between energy and
momentum. Thus, there is no propagating mode associated with
a timelike background vector. This property is an important difference between
the dimension-five model under consideration and MCFJ theory. The latter exhibits a
DR associated with a timelike background vector. However, this
timelike sector is plagued by consistency problems \cite{Adam1}.

\begin{figure}[t]
\centering
\includegraphics[scale=0.40]{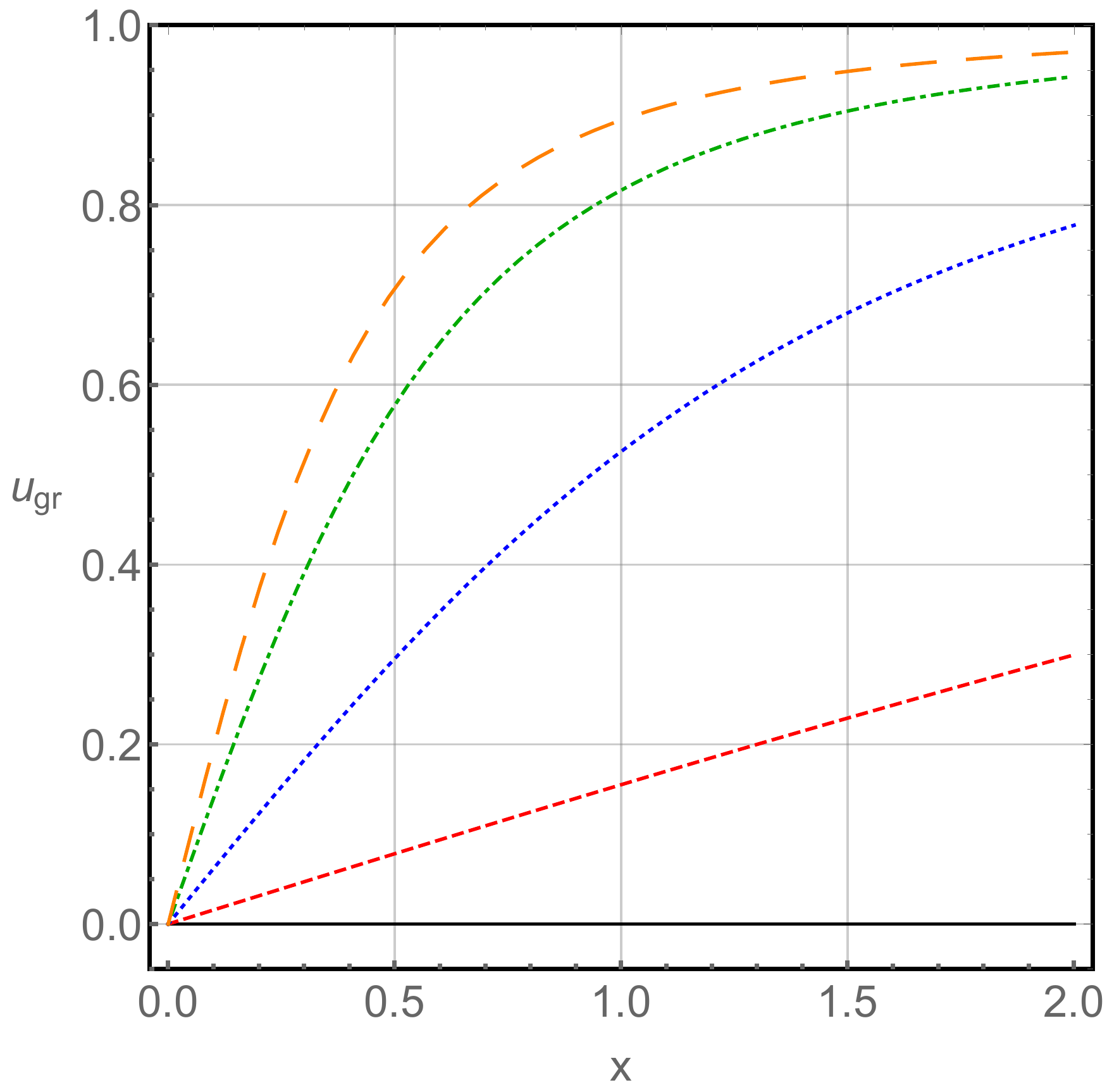}
\caption{Magnitude of the group velocity of Eq.~(\ref{eq:group-velocity}) for the spacelike case
with $\alpha=0$ (black, plain), $\alpha=\pi/40$ (red, dashed), $\alpha=\pi/10$ (blue, dotted),
$\alpha=\pi/4$ (green, dashed-dotted), and $\alpha=\pi/2$ (orange, long dashes).}%
\label{fig:group-velocity-spacelike}%
\end{figure}

Now, for a purely spacelike background, $D_{\gamma}=(0,\mathbf{D})_{\gamma}$,
the corresponding DR is%
\begin{equation}
p_{0}=\frac{1}{|\mathbf{D}|}\sqrt{\frac{1}{4}+\mathbf{D}^{2}\mathbf{p}%
^{2}-(\mathbf{D}\cdot\mathbf{p})^{2}}=\frac{1}{|\mathbf{D}|}\sqrt{\frac{1}%
{4}+|\mathbf{D}\times\mathbf{p}|^{2}}\,, \label{DR2b}%
\end{equation}
which can also be written as
\begin{subequations}
\begin{equation}
p_{0}=\frac{1}{|\mathbf{D}|}\sqrt{\frac{1}{4}+\mathbf{D}^{2}\mathbf{p}^{2}%
\sin^{2}\alpha}\,, \label{DR2c}%
\end{equation}
with the angle $\alpha$ enclosed by $\mathbf{D}$ and $\mathbf{p}$:
\begin{equation}
\mathbf{D}\cdot\mathbf{p}=|\mathbf{D}||\mathbf{p}|\cos\alpha\,.
\end{equation}
This is a DR that is compatible with the propagation of signals, whose
properties need to be examined. In the current section, we are especially
interested in classical causality that is characterized by the behavior of the
group and front velocity $\mathbf{u}_{\mathrm{gr}}$ and $u_{\mathrm{fr}}$,
respectively, where \cite{Brillouin:1960}
\end{subequations}
\begin{equation}
\mathbf{u}_{\mathrm{gr}}\equiv\frac{\partial p_{0}}{\partial\mathbf{p}%
}\,,\quad u_{\mathrm{fr}}\equiv\lim_{|\mathbf{p}|\mapsto\infty}\frac{p_{0}%
}{|\mathbf{p}|}\,. \label{GVFV}%
\end{equation}
Classical causality is established as long as both $u_{\mathrm{gr}}\equiv
|\mathbf{u}_{\mathrm{gr}}|\leq1$ and $u_{\mathrm{fr}}\leq1$. We now evaluate
these characteristic velocities for DR~(\ref{DR2b}). The front
velocity is
\begin{equation}
u_{\mathrm{fr}}=\lim_{|\mathbf{p}|\mapsto\infty}\sqrt{\frac{1}{4\mathbf{D}%
^{2}\mathbf{p}^{2}}+\sin^{2}\alpha}=\sin\alpha\,,
\end{equation}
as $\alpha\in\lbrack0,\pi]$.
Furthermore, we investigate the behavior of the group velocity:
\begin{equation}
\mathbf{u}_{\mathrm{gr}}=\frac{\mathbf{D}^{2}\mathbf{p}-\mathbf{D}%
(\mathbf{D}\cdot\mathbf{p})}{|\mathbf{D}|\sqrt{1/4+|\mathbf{D}\times
\mathbf{p}|^{2}}}\,,
\end{equation}
whose magnitude is
\begin{equation}
\label{eq:group-velocity}
u_{\mathrm{gr}}=\frac{|\mathbf{D}\times\mathbf{p}|}{\sqrt{1/4+|\mathbf{D}%
\times\mathbf{p}|^{2}}}=\frac{\sin\alpha}{\sqrt{1/(4x^{2})+\sin^{2}\alpha}}\,,
\end{equation}
where $x\equiv|\mathbf{D}||\mathbf{p}|$ is a dimensionless parameter. Large
momenta correspond to large $x$. Hence,
\begin{equation}
\lim_{|\mathbf{p}|\mapsto\infty}u_{\mathrm{gr}}=\lim_{x\mapsto\infty
}u_{\mathrm{gr}}=1\,,
\end{equation}
independently of the angle $\alpha$. As $u_{\mathrm{gr}}\leq1$ and
$u_{\mathrm{fr}}\leq1$, classical causality is established for the whole range
of LV coefficients and momenta.
Figure~\ref{fig:group-velocity-spacelike} presents the behavior of the magnitude
of the group velocity. The graph shows a monotonically increasing group velocity
that reaches the asymptotic value 1, which is a behavior in accordance with causality.
It shares this property with the spacelike sector of MCFJ theory where classical
causality is guaranteed, as well (see first paper of \cite{Adam2}). The mode obtained
here is exotic in the sense that it does not propagate when Lorentz violation
goes to zero. Hence, it does not approach the standard DR in this
limit. The mode must be understood as a high-energy effect that propagates in a
well-behaved manner for large momenta.

\subsection{Unitarity}
\label{sec:unitarity-first-model}

The next step is to study unitarity at tree-level, which is performed by means of the saturated
propagator \textit{SP }\cite{Veltman}. The latter is a scalar quantity that is
implemented by contracting the propagator with external physical currents $J^{\mu}$ as
follows:
\begin{equation}
\mathit{SP}\equiv J^{\mu}\mathrm{i}\Delta_{\mu\nu}J^{\nu}\,. \label{Sat2}%
\end{equation}
The current $J^{\mu}$ is assumed to be real and satisfies the conservation law $\partial_{\mu
}J^{\mu}=0$, which in momentum space reads $p_{\mu}J^{\mu}=0$. In accordance
with this method, unitarity is assured whenever the imaginary part of the
residue of the saturation \textit{SP} (evaluated at the poles of the
propagator) is nonnegative. A way of carrying out the calculation is to
determine the eigenvalues of the propagator matrix, evaluated at their own
poles with current conservation taken into account. For the propagator found
in Eq.~(\ref{eq:propagator-momentum-space}), the saturation is%
\begin{equation}
\mathit{SP}=-\mathrm{i}\left\{  \frac{J^{2}+4p^{2}(J\cdot D)^{2}}{p^{2}\left[
1+4(D^{2}p^{2}-(D\cdot p)^{2})\right]  }\right\}  \,,
\end{equation}
where $J^{\mu}L_{\mu\alpha}J^{\alpha}=-\mathrm{i}J^{\mu}\epsilon_{\mu
\alpha\kappa\lambda}J^{\alpha}D^{\kappa}p^{\lambda}=0$. Contracting the
propagator with two equal conserved currents corresponds to getting rid of all
gauge-dependent (and therefore, unhysical) contributions. These usually are
terms that involve four-momenta with free indices corresponding to the indices
of the Green's function. Although the term $L_{\mu\alpha}$ does not
have this structure, it vanishes when coupled to two external
conserved currents due to its antisymmetry. Hence, the
only Lorentz-violating contributions of the propagator that have an
impact on unitarity (based on our criterion) are the denominators and the symmetric term, $D_{\mu}%
D_{\nu}$, formed from a combination of two preferred spacetime directions.

Now, for the Maxwell pole, $p^{2}=0$, the residue of the saturation is%
\begin{equation}
\mathrm{Res}(\mathit{SP})|_{p^{2}=0}=-\mathrm{i}\left.  \left[  \frac{J^{2}%
}{1-4(D\cdot p)^{2}}\right]  \right\vert _{p^{2}=0}\,.
\end{equation}
At the pole $p^{2}=0$ it holds that $p_{0}^{2}=\mathbf{p}^{2}$. From the law
of current conservation we obtain $p_{0}J_{0}=\mathbf{p}\cdot\mathbf{J}$.
Therefore, we can cast the corresponding imaginary part into the form
\begin{equation}
\mathrm{Im}[\mathrm{Res}(\mathit{SP})|_{p^{2}=0}]=\frac{1}{1-4(D_{0}%
|\mathbf{p}|-\mathbf{D\cdot p})^{2}}\frac{|\mathbf{p}\times\mathbf{J}|^{2}%
}{|\mathbf{p}|^{2}}\geq 0\,. \label{SATpole1}%
\end{equation}
For any background, $D_{\mu}=(D_{0},\mathbf{D})_{\mu}$, the imaginary part of the saturation
(\ref{SATpole1}) is nonnegative for small momenta, but becomes negative as the
momentum increases. For a timelike configuration, $D_{\mu}=(D_{0}%
,0)_{\mu}$, and for a spacelike configuration, $D_{\mu}=(0,\mathbf{D})_{\mu}$,
the quantity (\ref{SATpole1}) becomes negative for $1/4<D_{0}^{2}|\mathbf{p}|^2$
and $1/4<(\mathbf{D\cdot p})^2$, respectively. So, unitarity is not assured
at the pole $p^{2}=0$ for all configurations possible.

For the second DR, Eq.~(\ref{DR2}), it only makes sense to examine the
spacelike configuration, $D_{\mu}=(0,\mathbf{D})_{\mu}$, where
\begin{equation}
p^2=\frac{1-4(\mathbf{D}\cdot\mathbf{p})^{2}}{4\mathbf{D}^{2}}\,.
\end{equation}
It is reasonable to write the saturation as
\begin{equation}
\mathit{SP}=-\mathrm{i}\left\{\frac{J^2+4p^2(\mathbf{D}\cdot\mathbf{J})^2}{p^2\left[1-4(\mathbf{D}^2p^2-(\mathbf{D}\cdot\mathbf{p})^2)\right]}\right\}
=\frac{\mathrm{i}}{p^2-\frac{1-4(\mathbf{D}\cdot\mathbf{p})^2}{4\mathbf{D}^2}}\left[\frac{J^2}{4\mathbf{D}^2p^2}+\frac{(\mathbf{D}\cdot\mathbf{J})^2}{\mathbf{D}^2}\right]\,.
\end{equation}
Its residue at this pole is
\begin{equation}
\label{eq:residue-model1-spacelike}
\mathrm{Res}(\mathit{SP})|_{p^{2}=\frac{1-4(\mathbf{D}\cdot\mathbf{p})^{2}}{4\mathbf{D}^{2}}}=\mathrm{i}\left[\frac{J^2}{1-4(\mathbf{D}\cdot\mathbf{p})^2}+\frac{(\mathbf{D}\cdot\mathbf{J})^2}{\mathbf{D}^2}\right]\,.
\end{equation}
Due to current conservation, $J_{0}=(\mathbf{p\cdot J})/p_{0}$, the four-current squared can be cast into
\begin{equation}
\label{eq:four-current-squared1}
J^{2}=-\frac{\mathbf{J}^{2}[1+4\left(\mathbf{D}\times\mathbf{p}\right) ^{2}]-4\mathbf{D}^{2}(\mathbf{J}\cdot\mathbf{p})^{2}}{1+4(\mathbf{D}\times\mathbf{p})^{2}}=-\frac{\mathbf{J}^2[1-4(\mathbf{D}\cdot\mathbf{p})^2]+4\mathbf{D}^2[\mathbf{J}^2\mathbf{p}^2-(\mathbf{J}\cdot\mathbf{p})^2]}{1+4(\mathbf{D}\times\mathbf{p})^2}\,.
\end{equation}
In the standard case, current conservation implies that $J^2<0$, i.e., any physical
four-current is spacelike. However, this property does not necessarily hold in Lorentz-violating
theories, anymore --- as shown by Eq.~(\ref{eq:four-current-squared1}).
Inserting this result into Eq.~(\ref{eq:residue-model1-spacelike}), leads to the residue
\begin{equation}
\label{eq:res4}
\mathrm{Res}(\mathit{SP})|_{p^{2}=\frac{1-4(\mathbf{D}\cdot\mathbf{p})^{2}}{4\mathbf{D}^{2}}}=\frac{\mathrm{i}}{\mathbf{D}^{2}}\left\{\frac{-4\mathbf{D}^{4}(\mathbf{J}%
	\times \mathbf{p})^{2}+(1-4(\mathbf{D\cdot p})^{2})\left[4(\mathbf{D\cdot J})^{2}(
	\mathbf{D}\times \mathbf{p})^{2}-(\mathbf{D}\times \mathbf{J}%
	)^{2}\right]}{\left(1-4(
	\mathbf{D\cdot p})^{2}\right)\left[
	1+4(\mathbf{D}\times \mathbf{p})^{2}\right]}\right\}\,.
\end{equation}%
There are configurations for which the imaginary part of the latter is positive. For example, we can choose
$\mathbf{p}$ parallel to $\mathbf{J}$, whereby Eq.~(\ref{eq:res4}) reduces to
\begin{equation}
\label{eq:res4b}
\mathrm{Res}(\mathit{SP})|_{p^{2}=\frac{1-4(\mathbf{D}\cdot\mathbf{p})^{2}}{4\mathbf{D}^{2}},\mathbf{p}\parallel\mathbf{J}}=\mathrm{i}\left[-\frac{\mathbf{J}^2}{1+4(\mathbf{D}\times\mathbf{p})^2}+\frac{(\mathbf{D}\cdot\mathbf{J})^2}{\mathbf{D}^2}\right]\,.
\end{equation}
The first (negative) term can be suppressed for large momenta as long as $\mathbf{D}\nparallel\mathbf{p}$. As the
second (positive) contribution does not depend on the momentum, the imaginary part of the residue can be positive
for large enough momenta. Hence, there are configurations of background field, large momenta, and external current
for which unitarity is valid. This behavior is in accordance with the previous interpretation that the mode is
exotic and must be interpreted as a high-energy effect. In a more specific way, if $\theta$ is the angle between
$\mathbf{D}$ and $\mathbf{J}$, we have $(\mathbf{D}\cdot\mathbf{J})^2=\mathbf{D}^{2}\mathbf{J}^{2}\cos^{2}\theta$ and
$(\mathbf{D}\times \mathbf{p})^{2}=\mathbf{D}^{2}\mathbf{p}^{2}\sin^{2}\theta$. Thus, the residue~(\ref{eq:res4b})
reads
\begin{equation}
\mathrm{Res}(\mathit{SP})|_{p^{2}=\frac{1-4(\mathbf{D}\cdot\mathbf{p})^{2}}{4\mathbf{D}^{2}},\mathbf{p}\parallel\mathbf{J}}=\mathrm{i}\left(-\frac{\mathbf{J}^2}{1+4\mathbf{D}^2\mathbf{p}^2\sin^2\theta}+\mathbf{J}^2\cos^2\theta\right)
=\mathrm{i}\mathbf{J}^2\sin^2\theta\left(\frac{4(\mathbf{D}\cdot\mathbf{p})^2-1}{4(\mathbf{D}\times\mathbf{p})^2+1}\right)\,,
\end{equation}
whose imaginary part is positive for
\begin{equation}
(\mathbf{D}\cdot\mathbf{p})^2>\frac{1}{4}\,.
\end{equation}
The latter condition assures unitarity. A more general case to be investigated is $\mathbf{D}\perp \mathbf{p}$
with $\mathbf{p}\nparallel\mathbf{J}$, for which $\mathbf{D}\cdot\mathbf{p}=0$ and
$(\mathbf{D}\times \mathbf{p})^{2}=\mathbf{D}^{2}\mathbf{p}^{2}$. To examine it, we exploit observer rotation
invariance and employ the following coordinate system:
\begin{equation}
\mathbf{D}=|\mathbf{D}|\hat{\mathbf{x}}\,,\quad \mathbf{p}=|\mathbf{p}|\hat{\mathbf{y}}\,,\quad \mathbf{J}\cdot\hat{\mathbf{z}}=|\mathbf{J}|\cos\alpha\,,
\end{equation}%
in which
\begin{subequations}
\begin{align}
\mathbf{D}\cdot \mathbf{J}&=|\mathbf{J}||\mathbf{D}|\sin\alpha\cos\phi\,, \\[2ex]
\mathbf{J}\cdot \mathbf{p}&=|\mathbf{J}||\mathbf{p}|\sin\alpha\sin\phi\,,
\end{align}
\end{subequations}
where $\phi$ is the angle between the $x$ axis and the projection of the vector $\mathbf{J}$ in the $x$-$y$ plane.\footnote{Note that there is no connection between the $x$ axis used here and the variable $x$ employed previously, e.g., in Eq.~(\ref{eq:group-velocity}).} For this
configuration, the residue~(\ref{eq:res4}) can be expressed as
\begin{equation}
\mathrm{Res}(\mathit{SP})|_{p^{2}=\frac{1-4(\mathbf{D}\cdot\mathbf{p})^{2}}{4\mathbf{D}^{2}}}=-\mathrm{i}\mathbf{J}^2\left(\frac{(1+4\mathbf{D}^2\mathbf{p}^2)\cos^2\alpha+\sin^2\alpha\sin^2\phi}{1+4\mathbf{D}^2\mathbf{p}^2}\right)\,.
\end{equation}
The imaginary part of the latter is always negative, which demonstrates unitarity violation for any choice of the momentum
and the angles.
A similar investigation for $\mathbf{J}\perp \mathbf{p}$ yields a residue whose imaginary part can be either positive or
negative, also providing unitarity violation in some situations.

To summarize, while causality is assured for any configuration of the purely spacelike background, unitarity
can hold, but does not do so necessarily. In the next section we will examine a more involved version of this first
higher-derivative model.

\section{Maxwell electrodynamics modified by a CPT-odd dimension-five higher-derivative term: a second model}
\label{sec:modification-sophisticated-model}

In the last section, we have examined a \textit{CPT}-odd, dimension-five,
nonminimal extension of the electromagnetic sector, specifically represented
by the CFJ-like Lagrange density
\begin{equation}
\frac{1}{2}\epsilon^{\kappa\lambda\mu\nu}A_{\lambda}(\hat{k}_{AF})_{\kappa}F_{\mu\nu}\,,
\end{equation}
where the background vector field $(\hat{k}_{AF})_{\kappa}$ was written according to Eq.~(\ref{eq:background-vector2A}). As a second possibility, we
propose the more sophisticated choice
\begin{equation}
(\hat{k}_{AF})_{\kappa}=(k_{AF}^{(5)})_{\kappa}^{\phantom{\kappa}\alpha_{1}\alpha_{2}}\partial_{\alpha_{1}}\partial
_{\alpha_{2}}=T_{\kappa}T^{\alpha_{1}}T^{\alpha_{2}}\partial_{\alpha_{1}%
}\partial_{\alpha_{2}}=T_{\kappa}(T\cdot\partial)^{2}\,,
\end{equation}
where $T_{\kappa}$ is a Lorentz-violating four-vector whose mass dimension is
\begin{equation}
[T_{\kappa}]=-1/3\,,
\end{equation}
or equivalently $[T_{\kappa}^3]=-1$. In contrast to the theory studied before, the
vectorlike background field is now also contracted with the additional derivatives.
This structure is supposed to render the properties of the current theory more involved
than those of the previously studied one. Modifying Maxwell's theory by including this term
into its Lagrange density, leads to a higher-derivative (dimension-five) anisotropic MCFJ-like
theory described by
\begin{equation}
\mathcal{L}=-\frac{1}{4}F_{\mu\nu}F^{\mu\nu}+\frac{1}{2}\epsilon
^{\kappa\lambda\mu\nu}A_{\lambda}T_{\kappa}(T\cdot\partial)
^{2}F_{\mu\nu}+\frac{1}{2\xi}(\partial_{\mu}A^{\mu})^{2}\,.
\label{L}%
\end{equation}
In App.~\ref{sec:mapping-nonminimal-sme} we establish the connection between the modifications
given by Eqs.~(\ref{Lg}), (\ref{L}) and the very general compilation of
Lorentz-violating contributions
listed in \cite{NMkost2}. Now, the latter Lagrangian can also be
written as
\begin{subequations}
\begin{equation}
\mathcal{L}=\frac{1}{2}A^{\mu}\Xi_{\mu\nu}A^{\nu},
\end{equation}
with the operator
\begin{equation}
\label{eq:operator-Xi}
\Xi_{\mu\nu}=\square\Theta_{\mu\nu}-2\tilde{L}_{\mu\nu}-\frac{1}{\xi}\square
\Omega_{\mu\nu}\,,
\end{equation}
\end{subequations}
sandwiched between two gauge fields. Note that Eq.~(\ref{L}) is directly linked
to the photon sector of Myers-Pospelov theory (cf.~the first paper of \cite{Myers1}). The latter
was initially constructed as a
prototype of higher-derivative Lorentz-violating theories and includes dimension-five modifications
of scalars, Dirac fermions, and photons. In \cite{Reyes:2010pv,Reyes:2013nca}, certain properties
of its photon sector are on the focus. Our Eq.~(\ref{L}) corresponds
to the theory studied in the latter references for the correspondence $T^{\mu}=g^{1/3}n^{\mu}$
(with their coupling constant $g$ and preferred direction $n^{\mu}$)
and the choice $\xi=-1$ of the gauge fixing parameter. The article \cite{Reyes:2010pv} is primarily
dedicated to studying classical causality. Unitarity is analyzed in \cite{Reyes:2013nca}
for a lightlike background vector $T_{\mu}$ based on the validity of the optical theorem at
tree-level. In what follows, we intend to discuss additional aspects of classical causality of
this particular theory. Furthermore, the forthcoming investigation of unitarity is new in the sense
that it relies on a different technique (investigation of the residues of the saturated
propagator) and it is carried out for different sectors of the theory (timelike and spacelike
preferred directions~$T_{\mu}$).

Now, comparing the operator $\Xi_{\mu\nu}$ of \eqref{eq:operator-Xi} to that of
Eq.~(\ref{eq:differential-operator-o}), it is no longer possible
to extract a d'Alembertian from it.
The projectors $\Theta_{\mu\nu}$, $\Omega_{\mu\nu}$ are given as before by Eq.~(\ref{Proj1})
and the Lorentz-violating antisymmetric operator now reads
\begin{equation}
\tilde{L}_{\mu\nu}=\epsilon_{\mu\kappa\lambda\nu}T^{\kappa}(T\cdot\partial)^{2}\partial^{\lambda}\,.
\end{equation}
To determine the propagator of the theory given by Eq.~(\ref{L}), we propose the following \textit{Ansatz}:
\begin{equation}
\Delta^{\mu}_{\phantom{\mu}\alpha}=a\Theta_{\phantom{\mu}\alpha}^{\mu}+b\tilde{L}_{\phantom{\mu}\alpha}^{\mu}%
+c\Omega_{\phantom{\mu}\alpha}^{\mu}+eT^{\mu}T_{\alpha}+f(T^{\mu}\partial_{\alpha}+T_{\alpha}\partial^{\mu})\,,
\end{equation}
that must satisfy the identity%
\begin{equation}
\Xi_{\nu\mu}\Delta_{\phantom{\mu}\alpha}^{\mu}=\eta_{\nu\alpha}\,.
\label{Id}%
\end{equation}
The algebra of the tensor operators $\Theta_{\nu\mu}$, $\tilde{L}_{\mu\nu}$, $\Omega_{\nu\mu}$, $T_{\nu}T_{\mu}$,
$T_{\nu}\partial_{\mu}$, and $T_{\mu}\partial_{\nu}$ is the same as that presented in Tab.~\ref{tab:closed-algebra}
with the replacement $D_{\mu}\rightarrow T_{\mu}$ to be performed. For brevity, we introduce
\begin{subequations}
\begin{align}
\rho&\equiv T^{\mu}\partial_{\mu}\,, \displaybreak[0]\\[2ex]
\tilde{\Gamma}_{\alpha\nu}&\equiv \tilde{L}_{\nu\mu}\tilde{L}_{\phantom{\mu}\alpha}^{\mu}=-\left[T_{\alpha}T_{\nu
}\square-(T_{\nu}\partial_{\alpha}+T_{\alpha}\partial_{\nu})
\rho+(\rho^{2}-T^{2}\square)\Theta_{\nu\alpha}+\Omega
_{\nu\alpha}\rho^{2}\right]\rho^{4}\,.
\end{align}
\end{subequations}
To find the scalar operators $a\dots f$, we start from the tensor equation
(\ref{Id}) and employ the algebra of the tensor operators as before. The
calculation is completely analogous. After performing some algebraic
simplifications, we obtain%
\begin{subequations}
\begin{equation}
a=\frac{\square}{\Lambda}\,,\quad b=\frac{2}{\square}a\,,\quad c=-\frac{\xi}{\square}-4\frac{\rho^{6}}{\square\Lambda}\,,\quad e=-4\frac{\rho^{4}}{\Lambda}\,,\quad f=4\frac{\rho^{5}}{\square\Lambda}\,,
\end{equation}
where%
\begin{equation}
\Lambda=\square^{2}+4(\rho^{2}-T^{2}\square)\rho^{4}\,.
\end{equation}
\end{subequations}
The form of the propagator in momentum space is:
\begin{align}
\Delta_{\mu\alpha}(p)&=\frac{-\mathrm{i}}{p^{2}\left\{p^{4}+4[
T^{2}p^{2}-(T\cdot p)^{2}](T\cdot p)^{4}\right\}}\left[p^{4}\eta_{\mu\alpha}-2\mathrm{i}p^{2}(T\cdot p)^2\varepsilon_{\mu\kappa\lambda\alpha}T^{\kappa}p^{\lambda}\right. \nonumber\\
&\phantom{{}={}}+\left.\left\{(1+\xi)\left[4(T\cdot p)^{6}-p^4\right]-4\xi T^{2}p^{2}(T\cdot p)
^{4}\right\} \frac{p_{\mu}p_{\alpha}}{p^{2}}\right. \nonumber\\
&\phantom{{}={}}\left.+\,4p^{2}(T\cdot p)^{4}T_{\mu}T_{\alpha}-4(T\cdot p)^{5}(T_{\mu}p_{\alpha}+T_{\alpha}p_{\mu})
\right]\,. \label{PropagatorNew}%
\end{align}
As before, the parts of the propagator independent of the gauge-fixing parameter are transversal, i.e.,
the result of Eq.~(\ref{eq:transversality-propagator}) can be carried over. Taking into
account the correspondence $T^{\mu}=g^{1/3}n^{\mu}$ and the choice $\xi=-1$,  we found that our
$\Delta_{\mu\alpha}/\mathrm{i}$ of Eq.~(\ref{PropagatorNew}) corresponds to Eq.~(72) of \cite{Reyes:2010pv}.

\subsection{Dispersion relations}
\label{sec:dispersion-relations-model-2}

As observed in the first model, the propagator poles provide two
dispersion equations, namely%
\begin{subequations}
\begin{align}
p^{2}&=0\,, \\
p^{4}+4[T^{2}p^{2}-(T\cdot p)^{2}](T\cdot p)^{4}&=0\,. \label{DRother}%
\end{align}
\end{subequations}
The first again corresponds to the usual Maxwell pole, while the second
involves LV modifications caused by the higher-derivative term.
We will analyze the second dispersion
equation for two configurations of the LV background, which exhibit preferred
directions in a spacetime modified by Lorentz violation. We need to notice that
the dispersion equation~(\ref{DRother}) is different from Eq.~(\ref{DR2}), as it still
makes sense when the LV coefficient vanishes, whereas expression~(\ref{DR2}) does
not. Hence, the classification ``exotic'' will be not used here. A DR
originating from Eq.~(\ref{DRother}) can be named spurious, when it exhibits
unphysical behavior, or unconventional, when it is well-behaved. For a purely timelike
background, $T_{\gamma}=(T_{0},0)_{\gamma}$, we have
\begin{equation}
\label{NDR1C}%
p_{0}^{\pm}=\frac{|\mathbf{p}|}{\sqrt{1\mp 2T_{0}^{3}|\mathbf{p}|}}\,,
\end{equation}
which is equal to Eq.~(19) in \cite{Reyes:2010pv}.
In contrast to the first model, there are now two distinct modified modes.
The notation $\oplus/\ominus$ refers to the mode with the plus and minus sign label, respectively.
The energy is well-defined for the mode $\ominus$, but not for the mode $\oplus$, for
which it is only meaningful as long as $|\mathbf{p}|<1/(2T_{0}^{3})$.
Furthermore, a sign change of the controlling coefficient $T_0$ simply interchanges the
two DRs. Hence, without restriction of generality, we assume a
nonnegative coefficient: $T_0\geq 0$.

On the other hand, for a purely spacelike background, $T_{\gamma}=(0,\mathbf{T})_{\gamma}$, the corresponding DR is
\begin{subequations}
\begin{equation}
\tilde{p}_{0}^{\pm}=|\mathbf{p}|\left(  1+2|\mathbf{T}|^{6}|\mathbf{p}|^{2}\cos^{4}\alpha
\pm2|\mathbf{T}|^{3}|\mathbf{p}|
\cos^{3}\alpha\sqrt{|\mathbf{T}|^{6}|\mathbf{p}|^{2}\cos^{2}\alpha+1}\right)^{1/2}\,, \label{NDR2}%
\end{equation}
with the angle $\alpha$ enclosed by $\mathbf{T}$ and $\mathbf{p}$:%
\begin{equation}
\mathbf{T}\cdot\mathbf{p}=|\mathbf{T}||\mathbf{p}|\cos\alpha\,.
\end{equation}
\end{subequations}
The latter is confirmed by Eq.~(29) in \cite{Reyes:2010pv} when the appropriate replacements are carried out.
Recall that $(\mathbf{T}\cdot\mathbf{p})^{3}$, $(\mathbf{T}\cdot\mathbf{p})^{4}\mathbf{T}^{2}$,
have mass dimension equal to 2, while $(\mathbf{T}\cdot\mathbf{p})^2\mathbf{T}^4$ is dimensionless.
It is possible to show that the energy (\ref{NDR2}) is real for any absolute value of the background
vector and direction relative to $\mathbf{p}$, that is, $\tilde{p}_{0}^{2}>0$. Besides,
the DR of the mode $\oplus$ is mapped to that of the mode $\ominus$ and vice versa
when the direction of $\mathbf{T}$ is reversed, i.e., when $\mathbf{T}\mapsto -\mathbf{T}$ or
$\alpha\mapsto \pi-\alpha$. When electromagnetic waves propagate along a direction perpendicular
to $\mathbf{T}$, Lorentz violation does not have any effect and the DR is standard.
In the suitable momentum range, Eqs.~(\ref{NDR1C})
and (\ref{NDR2}) represent DRs compatible with a propagation of signals.
\begin{figure}
\centering
\includegraphics[scale=0.4]{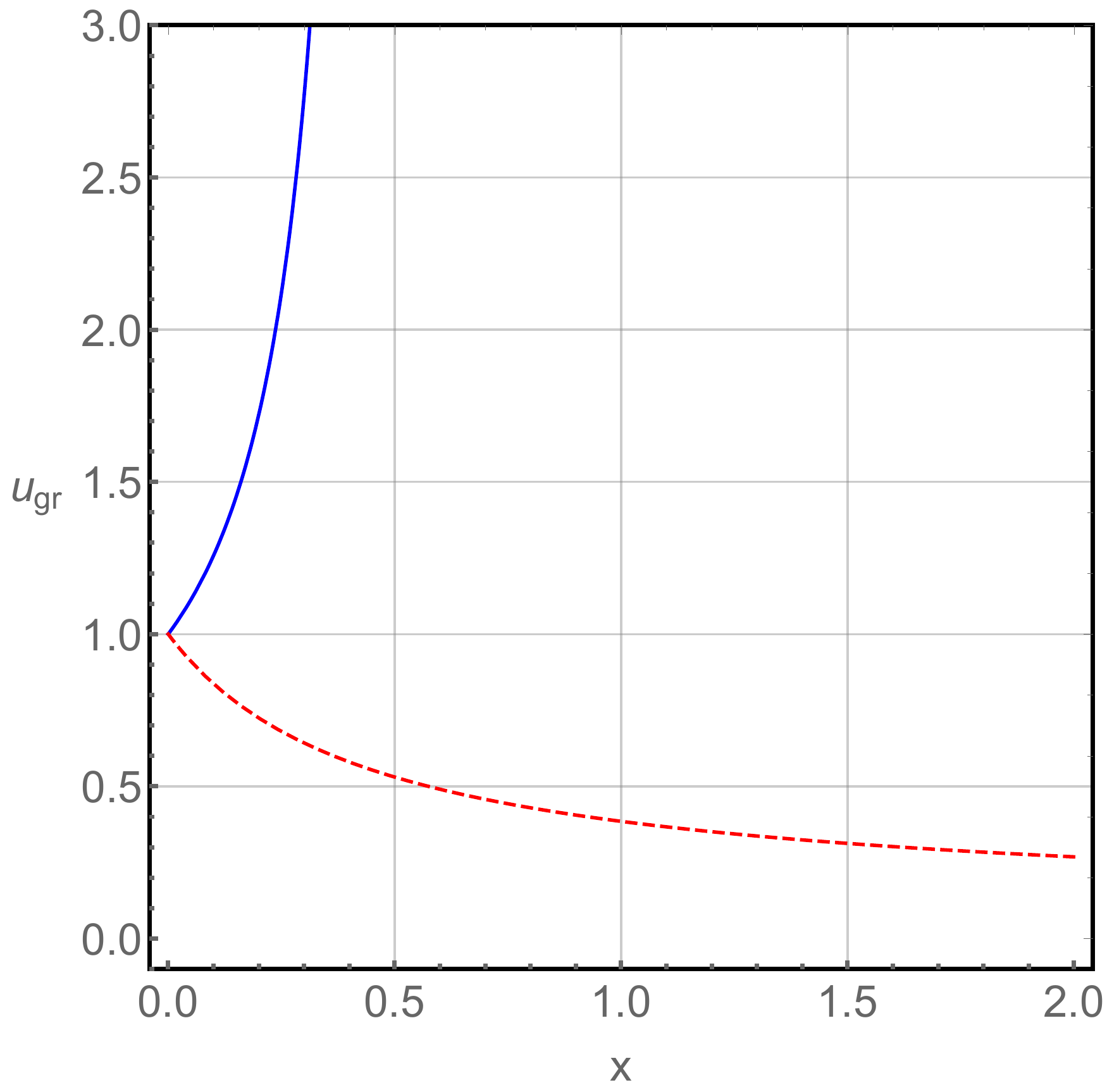}
\caption{Group velocity \eqref{eq:group-velocity-timelike-case} for the mode $\oplus$ (blue, plain) and
the mode $\ominus$ (red, dashed) as a function of $T_0^3|\mathbf{p}|\equiv x$.}
\label{fig:group-velocity-model2-timelike}
\end{figure}

As before, we investigate classical causality via the group and front velocity
$\mathbf{u}_{\mathrm{gr}}$ and $u_{\mathrm{fr}}$, respectively, defined in Eq.~(\ref{GVFV}).
We now evaluate these characteristic velocities for DRs originating from the dispersion
equation~(\ref{DRother}). On the one hand, we compute the front velocity for DR~(\ref{NDR1C}).
For the mode $\oplus$, the latter is not defined, as the energy becomes complex for momenta
beyond $1/(2T_0^3)$. For the mode $\ominus$, we obtain
\begin{equation}
u_{\mathrm{fr}}^{-}=\lim_{|\mathbf{p}|\mapsto\infty}\frac{p_0^{-}}{|\mathbf{p}|}=\lim_{|\mathbf{p}|\mapsto\infty}\frac{1}{\sqrt{1+2T_0^3|\mathbf{p}|}}=0\,.
\end{equation}%
On the other hand, we evaluate the group velocity:
\begin{equation}
\mathbf{u}_{\mathrm{gr}}^{\pm}=\frac{\partial p_0^{\pm}}{\partial\mathbf{p}}=\frac{1\mp T_0^3|\mathbf{p}|}{(1\mp 2T_0^3|\mathbf{p}|)^{3/2}}\frac{\mathbf{p}}{|\mathbf{p}|}\,,
\end{equation}
whose absolute values read (cf.~also Eq.~(21) in \cite{Reyes:2010pv})
\begin{equation}
\label{eq:group-velocity-timelike-case}
u_{\mathrm{gr}}^+=\left.\frac{1-T_0^3|\mathbf{p}|}{(1-2T_0^3|\mathbf{p}|)^{3/2}}\right|_{|\mathbf{p}|<1/(2T_0^3)}\,,\quad u_{\mathrm{gr}}^{-}=\frac{1+T_0^3|\mathbf{p}|}{(1+2T_0^3|\mathbf{p}|)^{3/2}}\,,
\end{equation}
where $u_{\mathrm{gr}}^+$ only makes sense for the range of momenta $|\mathbf{p}|<1/(2T_{0}^{3})$,
which is the same range that assures real energies for this mode. The behavior of $u_{\mathrm{gr}}^{\pm}$
is depicted in Fig.~\ref{fig:group-velocity-model2-timelike}. The large-momentum limit of
the group velocity for the mode $\ominus$ is simply
given by
\begin{equation}
\lim_{|\mathbf{p}|\mapsto\infty}u_{\mathrm{gr}}^-=0\,.
\end{equation}
The results for the front and group velocity indicate that signals do not propagate for large momenta.
Furthermore, $u_{\mathrm{gr}}^-$ is well-behaved for all momenta, whereas $u_{\mathrm{gr}}^+$
exhibits a singularity at $|\mathbf{p}|=1/(2T_0^3)$. For $|\mathbf{p}|>0$, classical
causality breaks down for this mode. Thus, the mode $\oplus$ must be considered as spurious. The mode $\ominus$ is
neither spurious nor exotic, but it is, indeed, an unconventional mode that does not propagate for
large momenta.

\begin{figure}[t]
\centering
\subfloat[]{\label{fig:group-velocity-model2-spacelike-positive}\includegraphics[scale=0.45]{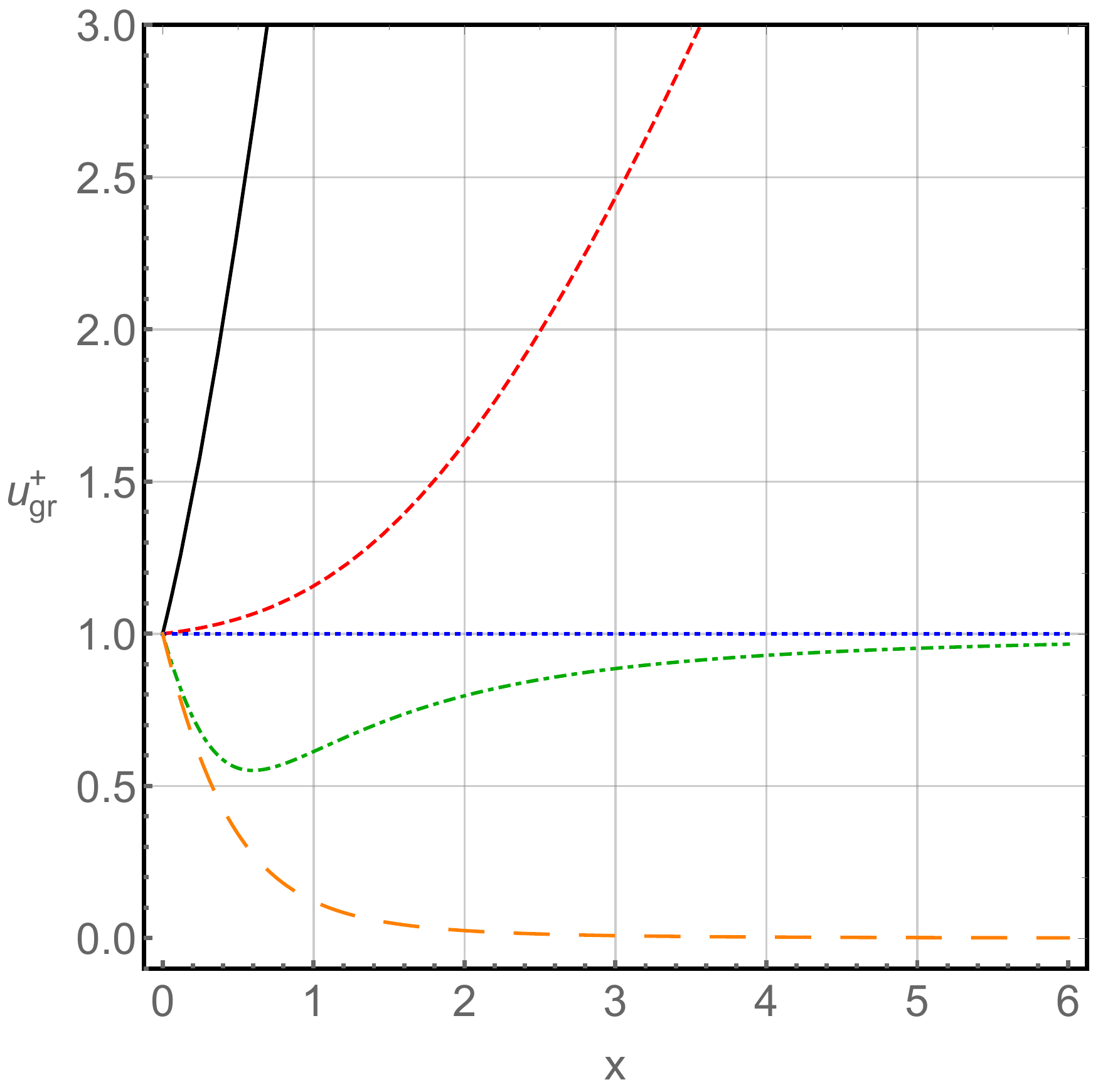}}\hspace{1cm}
\subfloat[]{\label{fig:group-velocity-model2-spacelike-negative}\includegraphics[scale=0.45]{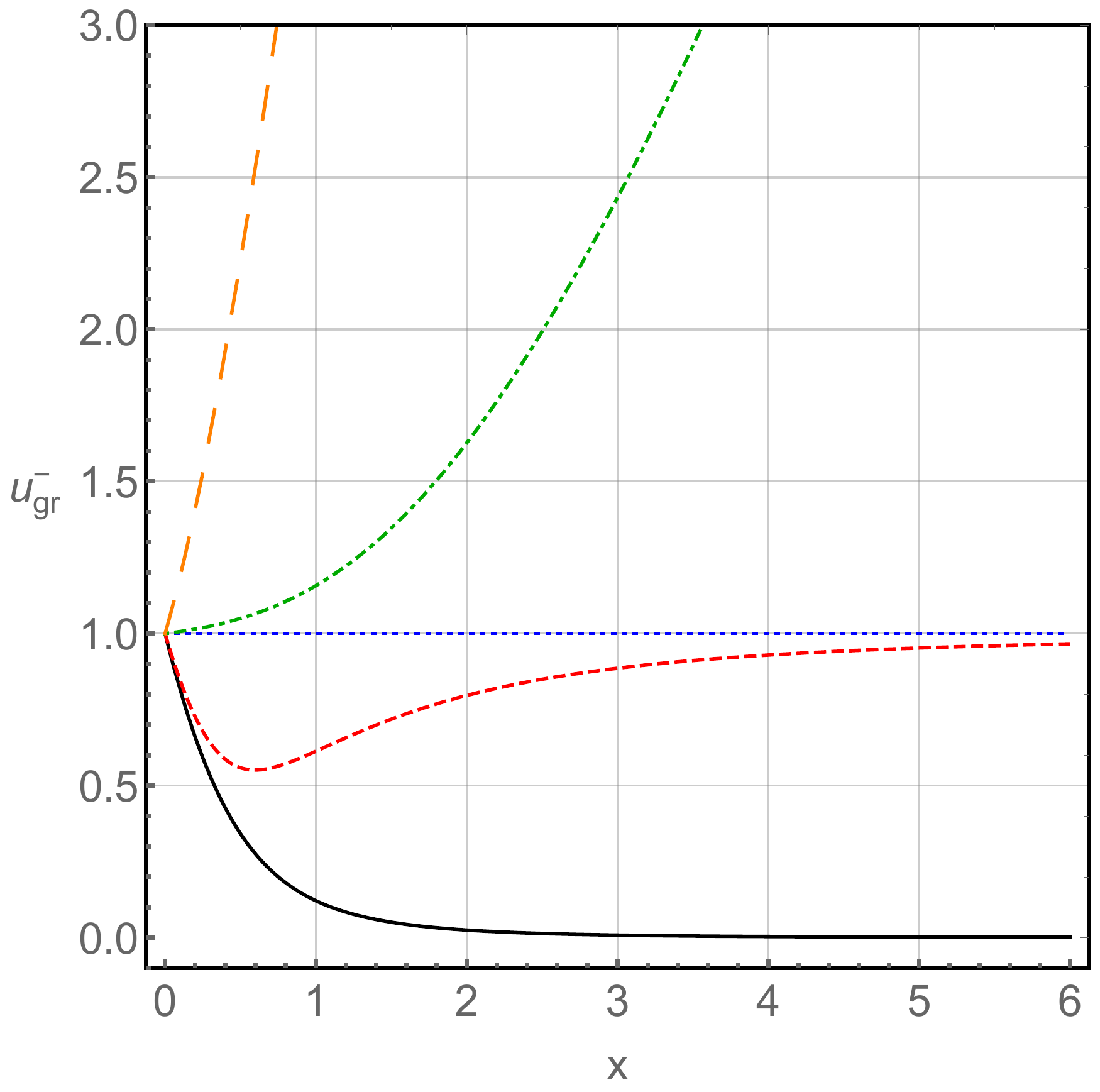}}
\caption{Group velocity of Eq.~(\ref{eq:group-velocity-spacelike}) for the mode $\oplus$ for $\alpha=0$ (black, plain), $\alpha=2\pi/5$ (red, dashed), $\alpha=\pi/2$
(blue, dotted), $\alpha=9\pi/10$ (green, dashed-dotted), and $\alpha=\pi$ (orange, long dashes)
\protect\subref{fig:group-velocity-model2-spacelike-positive}. Corresponding result for the
mode $\ominus$ for $\alpha=0$ (black, plain), $\alpha=\pi/10$ (red, dashed), $\alpha=\pi/2$
(blue, dotted), $\alpha=3\pi/5$ (green, dashed-dotted), and $\alpha=9\pi/10$ (orange, long dashes)
\protect\subref{fig:group-velocity-model2-spacelike-negative}.}
\label{GV2+}
\end{figure}

The next step is to investigate the properties of DR~(\ref{NDR2}). The associated front velocity is
calculated as
\begin{equation}
u_{\mathrm{fr}}^{\pm}=\lim_{|\mathbf{p}|\mapsto\infty}\frac{\tilde{p}_0^{\pm}}{|\mathbf{p}|}=\lim_{|\mathbf{p}|\mapsto\infty}\sqrt{1+2|\mathbf{T}|^{6}|\mathbf{p}|^{2}\cos^{4}%
\alpha\pm2|\mathbf{T}|^{6}|\mathbf{p}|^{2}\cos^{4}\alpha}\,,
\end{equation}
which provides%
\begin{equation}
\label{eq:front-velocity-spacelike-case}
u_{\mathrm{fr}}^{+}=\lim_{|\mathbf{p}|\mapsto\infty} 2\sqrt{|\mathbf{T}|^{6}|\mathbf{p}|^{2}\cos^{4}\alpha}=\infty\,,\quad u_{\mathrm{fr}}^{-}=1\,,
\end{equation}
independently of the choice of $\mathbf{T}$.
Hence, the front velocity is divergent for the mode $\oplus$, which again indicates
a breakdown of classical causality for this mode, whereby it is spurious. In contrast to that, the front velocity for
the mode $\ominus$ is well-behaved. What remains to be done, is to study the
properties of the group velocity:
\begin{equation}
\mathbf{u}_{\mathrm{gr}}^{\pm}=\frac{\partial\tilde{p}_0^{\pm}}{\partial\mathbf{p}}=\frac{\sqrt{1+(\mathbf{T}\cdot
		\mathbf{p})^{2}\mathbf{T}^{4}}\left[  \mathbf{p}+4(
	\mathbf{T}\cdot\mathbf{p})^{3}\mathbf{T}^{2}\mathbf{T}\right]\pm(
	\mathbf{T}\cdot\mathbf{p})^{2}\mathbf{T}\left[3+4(
	\mathbf{T}\cdot\mathbf{p})^{2}\mathbf{T}^{4}\right]}{\sqrt{1+(
		\mathbf{T}\cdot\mathbf{p})^{2}\mathbf{T}^{4}}\sqrt{\mathbf{p}%
		^{2}+2(\mathbf{T}\cdot\mathbf{p})^{4}\mathbf{T}^{2}\pm2(
		\mathbf{T}\cdot\mathbf{p})^{3}\sqrt{1+(\mathbf{T}%
			\cdot\mathbf{p})^{2}\mathbf{T}^{4}}}}\,.
\end{equation}
whose absolute value is
\begin{subequations}
\label{eq:group-velocity-spacelike}
\begin{equation}
u_{\mathrm{gr}}^{\pm}=\frac{\sqrt{2[4x^3\cos^4\alpha+3x\cos^2\alpha]^2+(x^2/4)\sin^2(2\alpha)+1\pm U(x)}}%
{\sqrt{1+x^{2}\cos^{2}\alpha}\sqrt{2x^{2}\cos^{4}\alpha+1\pm2x\cos^{3}%
\alpha\sqrt{1+x^{2}\cos^{2}\alpha}}}\,,
\end{equation}
where $|\mathbf{T}|^{3}|\mathbf{p}|\equiv x$ and%
\begin{equation}
U(x)=2x\cos^{3}\alpha[(2+4x^2\cos^2\alpha)^2-1]\sqrt{1+x^{2}\cos^{2}\alpha}\,.
\end{equation}
\end{subequations}
The plots shown in Fig.~\ref{GV2+} illustrate the behavior of the group velocity
for the unconventional modes $\oplus$ and $\ominus$ for different angles. For the mode $\oplus$,
the norm of the group velocity is equal 1 for $x=0$ and can be either larger or
smaller than 1 for $x>0$. For $\alpha\in [0,\pi/2)$, it steadily
becomes larger for $x>0$, which implies causality violation. For $\alpha\in (\pi/2,\pi)$,
it falls below 1 for $x>0$. In this regime, it has a minimum for a certain value of $x$ and
increases again whereupon it approaches 1 from below in the limit of large momenta. This behavior
is compatible with classical causality. For $\alpha=\pi$, the group velocity decreases
monotonically with $x$ until it reaches zero. Finally, the group velocity corresponds
to the standard result $u_{\mathrm{gr}}=1$ when $\alpha=\pi/2$, as the dispersion
relation is not modified in this case. The general behavior of the mode
$\ominus$ is analogous when $\alpha$ is replaced by $\pi-\alpha$.
Hence, the mode $\oplus$ cannot propagate for large momenta when the
momentum points in a direction opposite to $\mathbf{T}$ and a similar behavior
occurs for the mode $\ominus$ when the momentum is parallel to $\mathbf{T}$. Thus,
these results seem to show that the mode $\oplus$ preserves classical causality for
$\mathbf{T}\cdot\mathbf{p}\leq 0$, whereas $\ominus$ exhibits this property for
$\mathbf{T}\cdot\mathbf{p}\geq 0$. To conclude, the mode $\oplus$
is spurious for any choice of $\alpha$ due to Eq.~(\ref{eq:front-velocity-spacelike-case}),
whereas $\ominus$ is only spurious for $\mathbf{T}\cdot\mathbf{p}\leq 0$.

\subsection{Unitarity}

The analysis of unitarity at tree-level follows the same procedure applied in
Sec.~\ref{sec:unitarity-first-model}. For the propagator found in Eq.~(\ref{PropagatorNew}),
the saturation is%
\begin{equation}
\mathit{SP}=-\mathrm{i}\left\{\frac{p^{2}J^{2}+4(T\cdot p)^{4}(T\cdot J)^{2}}{p^{4}+4[T^{2}p^{2}-(T\cdot p)^{2}](T\cdot p)^{4}}\right\}\,.
\end{equation}
In contrast to the first model, the Maxwell pole cancels in the saturated propagator.
For the timelike configuration, $T_{\gamma}=(T_{0},0)_{\gamma}$, the DR
is given by Eq.~(\ref{NDR1C}). Then, the saturated propagator is
\begin{subequations}
\begin{align}
\mathit{SP}|_{\substack{\text{time-} \\ \text{like}}}&=-\mathrm{i}\left[\frac{p^{2}J^{2}+4T_{0}^{6}p_{0}^{4}J_{0}^{2}}{(1-4T_0^6\mathbf{p}^2)(p^{2}-p_{+}^{2})(p^{2}-p_{-}^{2})}\right]\,,
\end{align}
where
\begin{equation}
\label{eq:auxiliary-quantities-model2-timelike}
p_{\pm}^{2}\equiv p_{0\pm}^{2}-\mathbf{p}^2=\frac{\mathbf{p}^2}{1\mp 2T_0^3|\mathbf{p}|}-\mathbf{p}^2=\pm \frac{2T_0^3|\mathbf{p}|^3}{1\mp 2T_0^3|\mathbf{p}|}\,.
\end{equation}
\end{subequations}
We deduce that $p_+^2$ only makes sense for $|\mathbf{p}|<1/(2T_0^3)$ where it is larger than zero.
Besides, we observe that $p_-^2\leq 0$ for all values of $|\mathbf{p}|$. Apart from that, we have the
useful relation
\begin{equation}
p_+^2-p_-^2=\frac{4T_0^3|\mathbf{p}|^3}{1-4T_0^6\mathbf{p}^2}\,.
\end{equation}
The physical four-current squared can be conveniently expressed in the form
\begin{equation}
\label{eq:four-current-squared}
J^{2}=-\frac{1}{p_{0\pm}^{2}}\left[(\mathbf{J}\times\mathbf{p})^{2}+p_{\pm}^{2}\mathbf{J}^{2}\right]\,.
\end{equation}
Note that $p_-^2\leq 0$ according to Eq.~(\ref{eq:auxiliary-quantities-model2-timelike}),
which does not render the four-current squared negative for all configurations possible. Now, we use
Eq.~(\ref{eq:four-current-squared}) to write the residues for both poles as
\begin{equation}
\label{eq:resT01}
\mathrm{Res}(\mathit{SP})|_{p^{2}=p_{\pm}^{2}}=\pm\frac{\mathrm{i}}{p_{0\pm}^{2}}\left[\frac{p_{\pm}^{2}(\mathbf{J}\times\mathbf{p})^{2}+p_{\pm}^{4}\mathbf{J}^{2}-4T_{0}^{6}p_{0\pm}^{4}(\mathbf{J}\cdot\mathbf{p})^{2}}{4T_{0}^{3}|\mathbf{p}|^{3}}\right]\,.
\end{equation}
This expression is involved and can be better analyzed for some special configurations. For the particular
case $\mathbf{J}\parallel \mathbf{p}$, we obtain $(\mathbf{J}\times\mathbf{p})^{2}=0$ and
$\mathbf{J}\cdot\mathbf{p}=|\mathbf{J}||\mathbf{p}|$. Thus, the residue~(\ref{eq:resT01}) is%
\begin{equation}
\mathrm{Res}(\mathit{SP}|_{p^{2}=p_{\pm}^{2}})=\pm\mathrm{i}p_{0\pm}^{2}\mathbf{J}^{2}
\left(\frac{(p_{\pm}^{4}/p_{0\pm}^{4})-4T_{0}^{6}\mathbf{p}^{2}}%
{4T_{0}^{3}|\mathbf{p}|^{3}}\right)\,,
\end{equation}
which leads to a vanishing result when we take into account that
\begin{equation}
\frac{p_{\pm}^{2}}{p_{0\pm}^{2}}=\pm 2T_{0}^{3}|\mathbf{p}|\,.
\end{equation}
A vanishing residue means that the corresponding pole does not contribute to physical observables,
which is a situation compatible with unitarity. Another particular configuration
is $\mathbf{J}\perp\mathbf{p}$ for which $(\mathbf{J}\times\mathbf{p})^{2}=\mathbf{J}^{2}\mathbf{p}^{2}$ and
$\mathbf{J}\cdot\mathbf{p}=0$. In this case, the residue~(\ref{eq:resT01}) reduces to
\begin{equation}
\mathrm{Res}(\mathit{SP}|_{p^{2}=p_{\pm}^{2}})=\pm\mathrm{i}\frac{p_{\pm}^{2}}%
{p_{0\pm}^{2}}\mathbf{J}^{2}\left(\frac{\mathbf{p}^{2}+p_{\pm}^{2}}%
{4T_{0}^{3}|\mathbf{p}|^{3}}\right)\,,
\end{equation}
which can be simplified as
\begin{equation}
\mathrm{Res}(\mathit{SP}|_{p^{2}=p_{\pm}^{2}})=\mathrm{i}\frac{\mathbf{J}^{2}}%
{2}\left(\frac{\mathbf{1}}{1\mp2T_{0}^{3}\left\vert \mathbf{p}\right\vert
}\right)\,.
\end{equation}
The latter result confirms unitarity for both modes as far as the associated DRs are real,
which requires $|\mathbf{p}|<1/(2T_{0}^{3})$. Alternatively, the residue~(\ref{eq:resT01}) can be expressed
as
\begin{equation}
\label{eq:residue2}
\mathrm{Res}(\mathit{SP})|_{p^{2}=p_{\pm}^{2}}=\mp\mathrm{i}\frac{p_{\pm}^2J^2+(2T_0^3p_{0\pm}^2J_0)^2}{4(T_0|p^3|)^3}=\frac{\mathrm{i}}{2}\frac{(J^1)^2+(J^2)^2}{1\mp 2T_0^3|p^3|}\,,
\end{equation}
which holds in an observer frame where the three-momentum points along the third axis: $\mathbf{p}=(0,0,p^3)$.
Due to observer Lorentz invariance and isotropy of the theory considered, such a choice does not
restrict generality, confirming the particular results obtained for $\mathbf{J}\parallel\mathbf{p}$ and
$\mathbf{J}\perp\mathbf{p}$. For the mode $\oplus$, the imaginary part of the residue is nonnegative for
$|\mathbf{p}|<1/(2T_0^3)$, assuring unitarity in the momentum range in which DR~(\ref{NDR1C}) is real.
But unitarity of this mode is violated for $|\mathbf{p}|\geq1/(2T_0^3)$, when the energy
associated becomes complex. This is an expected breakdown, therefore.

The imaginary part of the residue is manifestly nonnegative for the
mode $\ominus$, ensuring unitarity for the full momentum range. These properties are in accordance with
the observations made in Sec.~\ref{sec:dispersion-relations-model-2} and Fig.~\ref{fig:group-velocity-model2-timelike}.
In contrast to $\oplus$, the mode $\ominus$ is well-behaved with respect to both classical causality and unitarity at tree-level.
However, it is interesting to mention that a breakdown of classical causality does not necessarily imply unitarity violation, as can
be seen for the spurious mode $\oplus$ where $|\mathbf{p}|\in [0,1/(2T_0^3))$.

For the purely spacelike choice, $T_{\gamma}=(0,\mathbf{T})_{\gamma}$, the saturated propagator is
\begin{subequations}
\begin{equation}
\mathit{SP}|_{\substack{\text{space-} \\ \text{like}}}=-\mathrm{i}\left[\frac{p^{2}J^{2}+4(\mathbf{T}\cdot\mathbf{p})^{4}(\mathbf{T}\cdot\mathbf{J})^{2}}{(p^{2}-\tilde{p}_{+}^{2})(p^{2}-\tilde{p}_{-}^{2})}\right]\,,
\end{equation}
where in the same manner as before, we introduce the auxiliary quantities
\begin{equation}
\tilde{p}_{\pm}^{2}\equiv \tilde{p}_{0\pm}^{2}-\mathbf{p}^2=2(\mathbf{T}\cdot\mathbf{p})^{3}\left[(\mathbf{T}\cdot\mathbf{p})\mathbf{T}^{2}\pm\sqrt{1+(\mathbf{T}\cdot\mathbf{p}) ^{2}\mathbf{T}^{4}}\right]\,.
\end{equation}
\end{subequations}
The four-current squared can be written as in Eq.~%
\eqref{eq:four-current-squared} with the replacements $p_{0\pm }\mapsto
\tilde{p}_{0\pm }$ and $p_{\pm }\mapsto \tilde{p}_{\pm }$. We observe that $%
\tilde{p}_{+}^{2}\leq 0$ for $\mathbf{T}\cdot \mathbf{p}\leq 0$ and $\tilde{p%
}_{-}^{2}<0$ for $\mathbf{T}\cdot \mathbf{p}\geq 0$. Therefore, the standard
condition $J^{2}<0$ for a conserved current does again not necessarily hold
in the Lorentz-violating context. Based on this result, we write the
residues of the two poles in the form
\begin{equation}
\mathrm{Res}(\mathit{SP})|_{p^{2}=\tilde{p}_{\pm }^{2}}=\pm \frac{\mathrm{i}%
}{\tilde{p}_{0\pm }^{2}}\left[ \frac{\tilde{p}_{\pm }^{2}(\mathbf{J}\times
	\mathbf{p})^{2}+\tilde{p}_{\pm }^{4}\mathbf{J}^{2}-4\tilde{p}_{0\pm }^{2}(%
	\mathbf{T}\cdot \mathbf{p})^{4}(\mathbf{T}\cdot \mathbf{J})^{2}}{4(\mathbf{T}%
	\cdot \mathbf{p})^{3}\sqrt{1+(\mathbf{T}\cdot \mathbf{p})^{2}\mathbf{T}^{4}}}%
\right] \,.
\end{equation}%
\begin{figure}[b]
\centering
\subfloat[]{\label{function-positive}\includegraphics[scale=0.5]{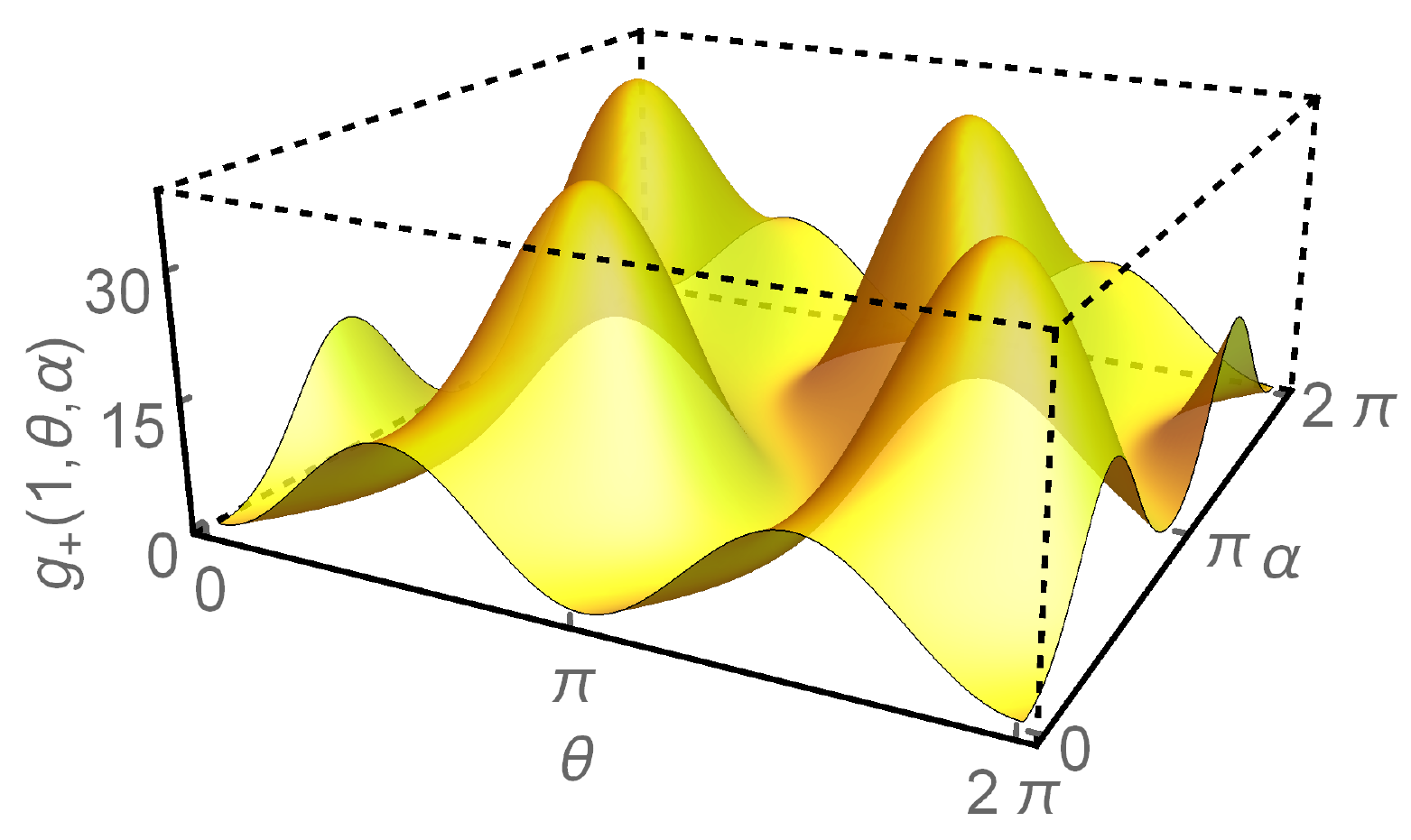}}\hspace{1cm}
\subfloat[]{\label{function-negative}\includegraphics[scale=0.5]{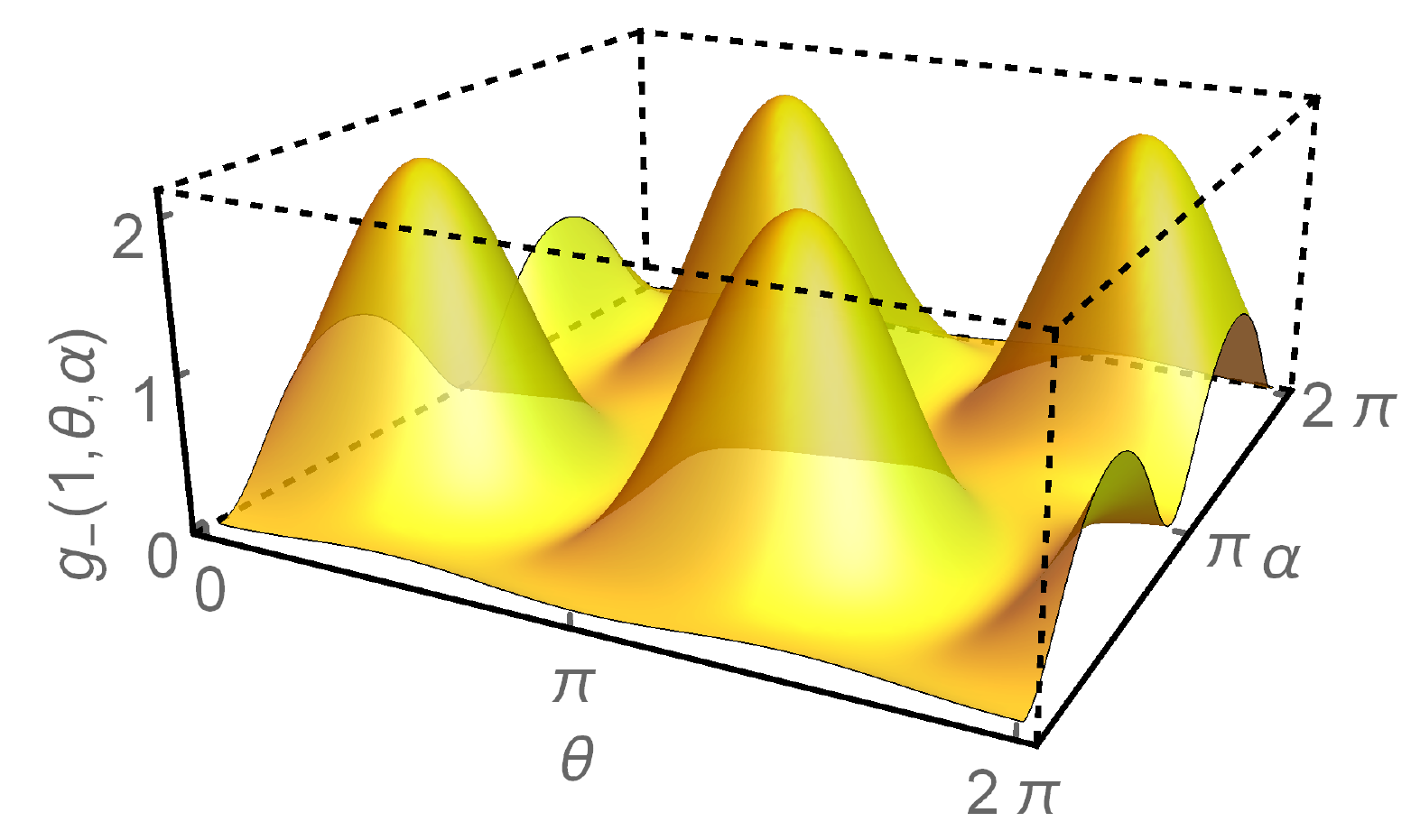}}
\caption{Functions $g_{\pm}(1,\theta,\alpha)$ of Eq.~(\ref{eq:model2-spacelike-numerator}) for the mode $\oplus$ \protect\subref{function-positive} and $\ominus$ \protect\subref{function-negative}.}
\label{fig:functions-g-plus-minus}
\end{figure}
The latter can also be cast into
\begin{subequations}
	\begin{equation}
	\mathrm{Res}(\mathit{SP})|_{p^{2}=\tilde{p}_{\pm }^{2}}=\mathrm{i}\left(
	\frac{N_{\pm}}{2\tilde{p}_{0\pm }^{2}\sqrt{1+(\mathbf{T}\cdot \mathbf{p})^{2}%
			\mathbf{T}^{4}}}\right)\,,  \label{eq:residue-model2-spacelike-case}
	\end{equation}%
	with
	\begin{align}
	N_{\pm}& =\mp\left\{ 2(\mathbf{T}\cdot \mathbf{p})(\mathbf{T}\cdot \mathbf{J})^{2}%
	\mathbf{p}^{2}+4(\mathbf{T}\cdot \mathbf{p})^{4}(\mathbf{T}\cdot \mathbf{J}%
	)^{2}\Upsilon _{\pm }-2\mathbf{J}^{2}\Upsilon _{\pm }^{2}(\mathbf{T}\cdot
	\mathbf{p})^{3}-\Upsilon _{\pm }(\mathbf{J}\times \mathbf{p})^{2}\right\} ,
	\\[2ex]
	N_{\pm}& =\mp\left\{ 2(\mathbf{T}\cdot \mathbf{p})\left[ \mathbf{p}^{2}(\mathbf{T}%
	\cdot \mathbf{J})^{2}-\mathbf{J}^{2}(\mathbf{T}\cdot \mathbf{p})^{2}\right]
	-\Upsilon _{\pm }\left[ 4(\mathbf{T}\cdot \mathbf{p})^{4}(\mathbf{J}\times
	\mathbf{T})^{2}+(\mathbf{J}\times \mathbf{p})^{2}\right] \right\} \,,
	\end{align}%
	and%
	\begin{equation}
	\Upsilon _{\pm }=(\mathbf{T}\cdot \mathbf{p})\mathbf{T}^{2}\pm \sqrt{1+(%
		\mathbf{T}\cdot \mathbf{p})^{2}\mathbf{T}^{4}}\,.
	\label{eq:model2-quantities-upsilon}
	\end{equation}%
	Although the residue of Eq.~(\ref{eq:residue-model2-spacelike-case}) does
	not look that involved, an analysis of this expression turned out to be
	challenging. Based on the new quantities of Eq.~(\ref%
	{eq:model2-quantities-upsilon}), the modified DRs can be
	expressed in a short manner via
\end{subequations}
\begin{equation}
\tilde{p}_{0}^{2}=\mathbf{p}^{2}+2(\mathbf{T}\cdot \mathbf{p})^{3}\Upsilon
_{\pm }\,.
\end{equation}%
A further useful relation in this context is
\begin{equation}
\label{eq:DRSP4}
\Upsilon _{\pm }^{2}=1+2(\mathbf{T}\cdot \mathbf{p})\mathbf{T}^{2}\left[ (%
\mathbf{T}\cdot \mathbf{p})\mathbf{T}^{2}\pm \sqrt{1+(\mathbf{T}\cdot
	\mathbf{p})^{2}\mathbf{T}^{4}}\right] =1+2(\mathbf{T}\cdot \mathbf{p})%
\mathbf{T}^{2}\Upsilon _{\pm }\,.
\end{equation}
According to our criterion, unitarity is guaranteed as long as $N_{\pm}\geq 0$.
The second contribution of $N_{\pm}$ is manifestly nonnegative due to $\pm\Upsilon_{\pm}>0$. However, the first is not. For the configurations $\mathbf{J}\parallel\mathbf{p}$ and $\mathbf{T}\perp\mathbf{p}$, the first term simply vanishes, i.e., unitarity can be demonstrated quickly for these special cases. In particular, the imaginary part of the residue for $\mathbf{J}\parallel\mathbf{p}$ is
\begin{equation}
\label{eq:residue-model2-spacelike-case4}
\mathrm{Im}[\mathrm{Res}(\mathit{SP})|_{p^{2}=\tilde{p}_{\pm}^{2}}]=\pm\frac{2\Upsilon_{\pm}(\mathbf{T}\cdot\mathbf{p})^4(\mathbf{J}\times\mathbf{T})^2}{\tilde{p}_{0\pm}^2\sqrt{1+(\mathbf{T}\cdot\mathbf{p})^2\mathbf{T}^4}}\geq 0\,.
\end{equation}
Specifically for the situation $\mathbf{T}\perp\mathbf{p}$, DR~(\ref{eq:DRSP4}) recovers the usual Maxwell DR and $\Upsilon_{\pm}=1$, so that the residue simply yields
\begin{equation}
\mathrm{Im}[\mathrm{Res}(\mathit{SP})|_{p^2=0}]=\frac{|\mathbf{J}\times\mathbf{p}|^{2}}{2\mathbf{p}^{2}}\geq 0\,. \label{SATpole2}%
\end{equation}
The origin of the factor of 2 in the denominator is due to the existence of two distinct DRs for
nonzero Lorentz violation and configurations other than $\mathbf{T}\perp\mathbf{p}$.
When both merge for vanishing Lorentz violation, each contributes the above value to the residue providing its standard
result.
\begin{figure}
\centering
\subfloat[]{\label{function-positive-zeros}\includegraphics[scale=0.5]{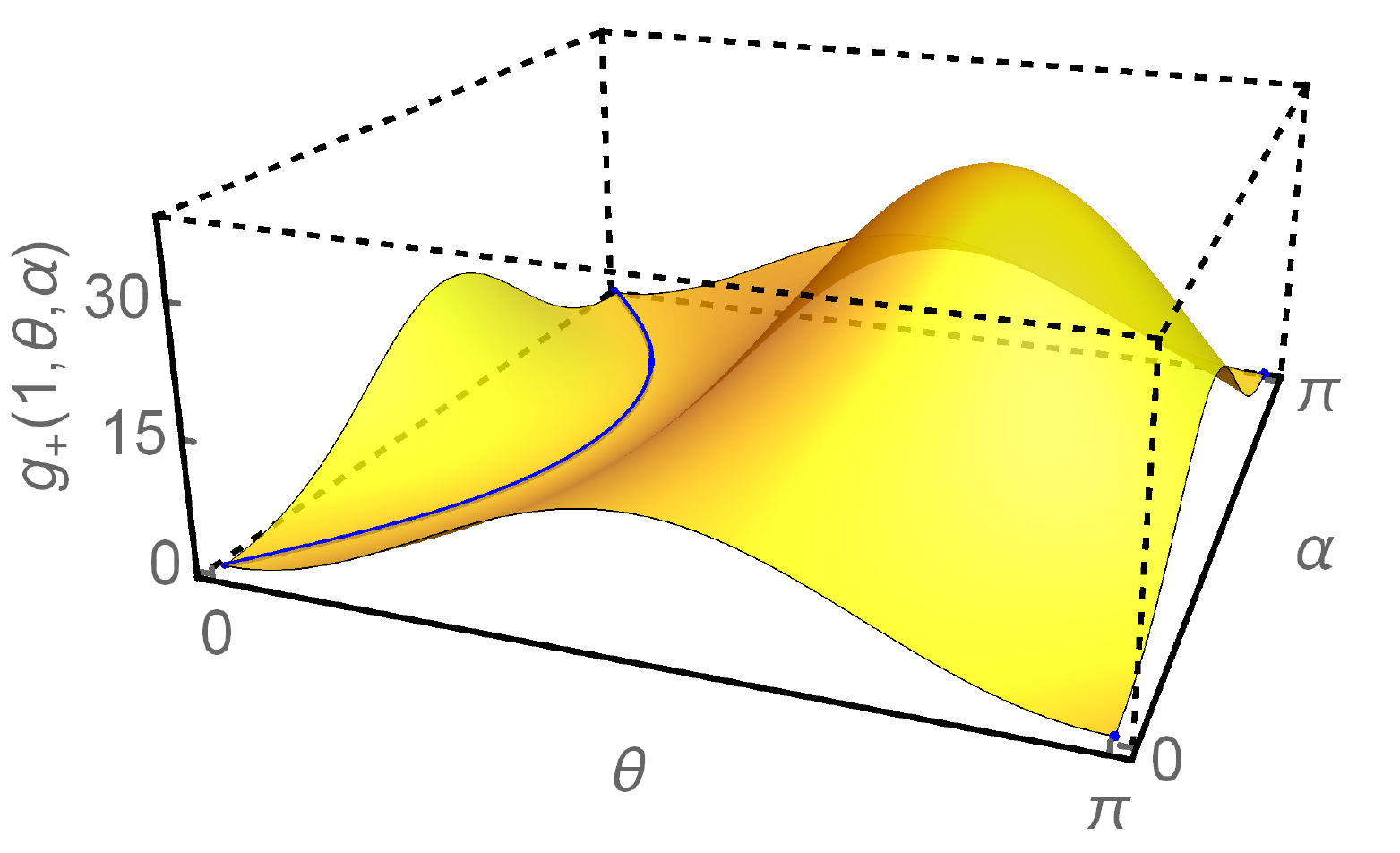}}\hspace{1cm}
\subfloat[]{\label{function-negative-zeros}\includegraphics[scale=0.5]{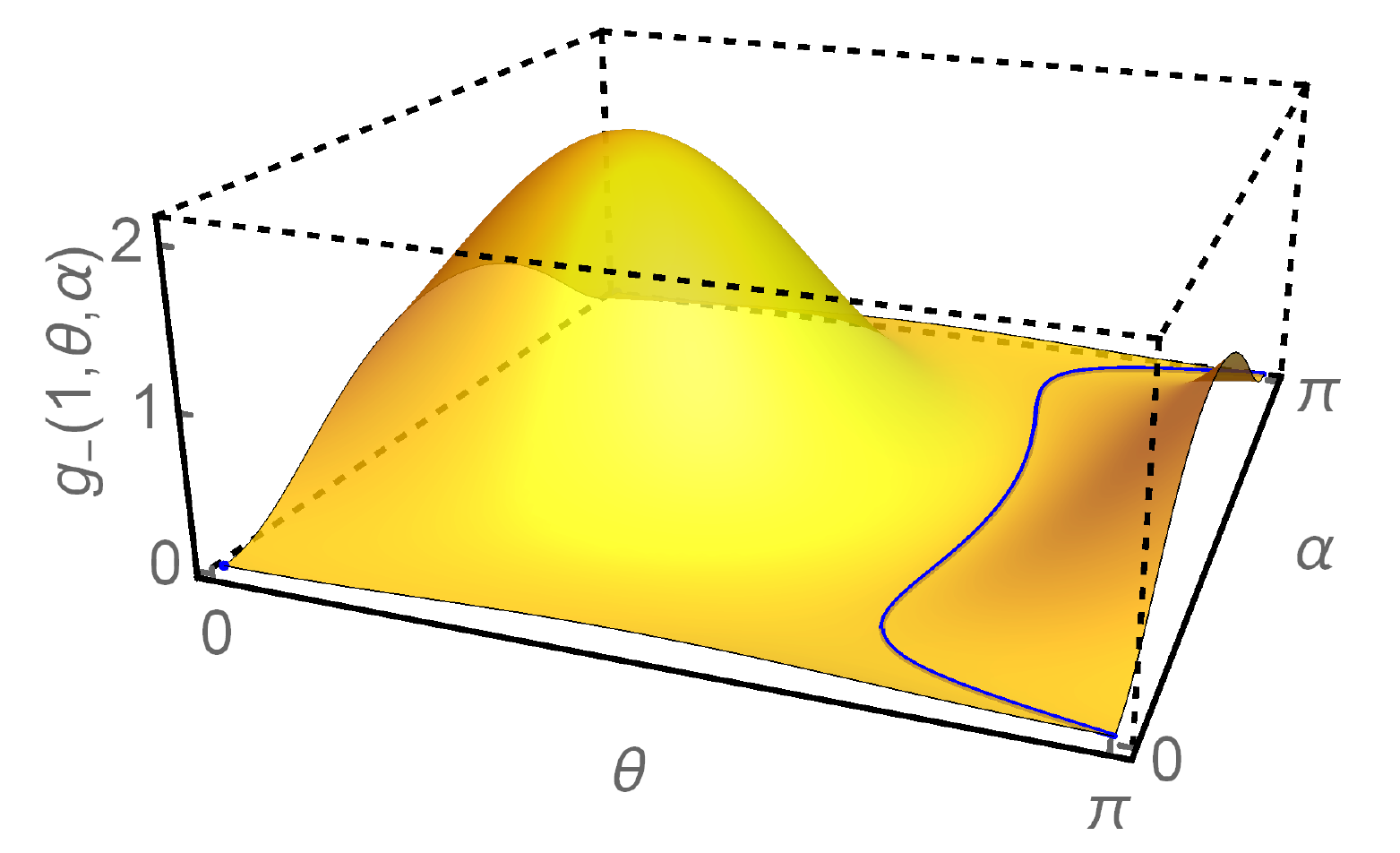}}
\caption{Functions $g_{\pm}(1,\theta,\alpha)$ of Eq.~(\ref{eq:model2-spacelike-numerator}) in the regime $\alpha$, $\theta\in [0,\pi]$ for the mode $\oplus$ \protect\subref{function-positive-zeros} and $\ominus$ \protect\subref{function-negative-zeros}. The zeros obtained numerically lie along the plain, blue lines.}
\label{fig:functions-g-plus-minus-zeros}
\end{figure}

In contrast, for most other choices, an evaluation of the inequality $N_{\pm}\geq 0$ seems to be highly involved. What can be done, is to employ observer rotational invariance and to choose a coordinate system such that the momentum points along the third axis and $\mathbf{T}$, $\mathbf{J}$ lie in the plane spanned by the first and third axes. So we consider
\begin{equation}
\mathbf{p}=\begin{pmatrix}
0 \\
0 \\
|\mathbf{p}| \\
\end{pmatrix}\,,\quad \mathbf{T}=|\mathbf{T}|\begin{pmatrix}
\sin\alpha \\
0 \\
\cos\alpha \\
\end{pmatrix}\,,\quad \mathbf{J}=|\mathbf{J}|\begin{pmatrix}
\sin\theta \\
0 \\
\cos\theta \\
\end{pmatrix}\,,
\end{equation}
with $\alpha\in [0,2\pi]$ and $\theta\in [0,2\pi]$. Inserting these representations into the left-hand side
of $N_{\pm}\geq 0$, we obtain
\begin{subequations}
\begin{align}
\frac{N_{\pm}}{\mathbf{J}^2\mathbf{p}^2}&\equiv g_{\pm}(\xi,\theta,\alpha)\geq 0\,, \\[2ex]
\label{eq:model2-spacelike-numerator}
g_{\pm}(\xi,\theta,\alpha)&=\pm\,4\xi^3(1+\sin^2\alpha)(\sin\theta-\sin\alpha\cos\theta)^2 \notag \\
&\phantom{{}={}}+\left[4\xi^2(\sin\theta-\sin\alpha\cos\theta)^2+\sin^2\theta\right]\sqrt{1+(1+\sin^2\alpha)^2\xi^2} \notag \\
&\phantom{{}={}}\pm\xi\left[(2+\cos^2\alpha)\sin^2\theta-2\sin\alpha\sin(2\theta)\right]\,,
\end{align}
with
\begin{equation}
\xi\equiv |\mathbf{p}||\mathbf{T}|^3\,.
\end{equation}
\end{subequations}
The left-hand side of the new inequality are functions of $\xi\geq 0$ and the two angles $\alpha$, $\theta$. It is challenging to prove that $g_{\pm}(\xi,\theta,\alpha)$ is nonnegative for general $\xi$ and angles. Plots of these functions for a given choice of $\xi$ are presented in Fig.~\ref{fig:functions-g-plus-minus}. According to the plots, both functions most probably do not provide negative values. Numerical investigations allow for determining the zeros of $g_{\pm}(1,\theta,\alpha)$; cf.~Fig.~\ref{fig:functions-g-plus-minus-zeros}. There are no isolated zeros, but they lie along certain lines. Evaluating the first and second derivatives at these points, indicates that they are local minima. Minima other than those were not found. Hence, there are strong numerical indications that $g_{\pm}(1,\theta,\alpha)$ are nonnegative. Analog results can be obtained for $\xi\neq 1$, ensuring unitarity for such more general configurations of the purely spacelike case.

\section{Conclusions and final remarks}
\label{sec:conclusions}

We analyzed Maxwell's electrodynamics modified by dimension-five and
\textit{CPT}-odd higher-derivative terms that are part of the photon sector
of the nonminimal Standard-Model Extension. The first dimension-five term to
be addressed was a kind of higher-derivative CFJ-like contribution.
The propagator of the theory was computed by means of the algebra for the
tensor operators involved. Based on this result,
modifications of the properties of wave propagation
due to the presence Lorentz violation were investigated.

The analysis of the dispersion relations obtained from the
propagator poles revealed that signals do not propagate at all for the purely
timelike sector. The modes of the purely spacelike sector decouple from the
theory for low energies and only propagate in the high-energy regime. For
this sector, classical causality is preserved for any choice of background
coefficients. Furthermore, unitarity at tree-level was examined
by contracting the propagator with conserved currents and studying the pole
structure of the resulting expression. It was found that unitarity
can be preserved in some special cases. In general, however, the dispersion relations
describe nonunitary modes.  This propagator possesses some analogy with
that of MCFJ theory, which was shown in App.~\ref{sec:comparison-propagator-mcfj-theory}.
But the dispersion relations and the related physics of these two models differ
from each other.

The second dimension-five term examined was an anisotropic
higher-derivative CFJ-like contribution that can be identified with the photon
sector of Myers-Pospelov theory. The propagator
was again computed and the dispersion relations were obtained from it
for a purely timelike and a purely spacelike background vector field. For the timelike background, one mode was found to be unphysical, whereas the second generally obeys
classical causality and unitarity, as shown in Eq.~(\ref{eq:residue2}). For the spacelike configuration, there are
two modes, as well. One mode is causal when the propagation
direction encloses an angle $\alpha\in [0,\pi/2]$ with the spacelike preferred
direction and noncausal for complementary angles. The other mode is noncausal for
all angles $\alpha$. However, numerical investigations indicate that unitarity
is preserved for a large subset of configurations within this scenario (in spite of causality
violation).

The results of this paper complement our findings for nonminimal
\textit{CPT}-even extensions of the electromagnetic sector considered in \cite{Leticia1}.
In general, the studies performed for the \textit{CPT}-odd extensions were easier from
a technical point of view in comparison to their \textit{CPT}-even counterparts.
Furthermore, our results reveal that a breakdown of classical
causality does not necessarily imply unitarity violation at tree-level. Indeed,
with regards to our results and those of \cite{Reyes:2013nca}, we can safely conclude that the
photon sector of Myers-Pospelov theory is well-behaved with respect to unitarity for a large
number of parameter choices --- at least based on the criteria studied in the current work and
in \cite{Reyes:2013nca}. This is the case even for the modes that violate causality.
It is also known that a violation of classical causality does not always
yield a violation of causality at the microscopic level, as was demonstrated for particular
minimal models; cf.~\cite{Klinkhamer:2010zs}. Therefore, noncausal modes in this sense can still
be of interest.

On the other hand, the fate of nonunitary modes is uncertain, as long as
contributions of even higher mass dimensions that may cure this malign behavior
are not taken into account. Another point connected with quantization and
unitarity of this theory is how to suitably fix the gauge for addressing the additional degrees of
freedom that usually appear in higher-derivative theories. Such cases can require a nonstandard
gauge fixing condition. Another question to be tackled is if there is an analog of the
Lee-Wick mechanism that allows us to preserve unitarity by decoupling the ghost modes and
removing the negative-norm states from the asymptotic Hilbert space, even in the sectors
plagued by nonunitarity behavior. One possibility of exploring this issue is to couple
the higher-derivative electrodynamics to fermions and to use the optical theorem, as carried out
in Ref.~\cite{Reyes:2017pus}.

CJF theory, which is {\em CPT}-odd and of dimension 3, is considered as a prototype Lorentz-violating
modification of the photon sector that has been of great interest over the past 30 years. As its
properties have already been studied in great detail, it was a natural step to look at its higher-dimensional
versions of which the photon sector of Myers-Pospelov theory is one possible extension. We argued at the beginning that
dimension-three modifications are suppressed for high energies, whereas the dimension-five terms studied
here gain importance in this regime. The results of the paper indicate which sectors of the theory
should be disregarded and which are suitable for quantization and phenomenological studies. Due to
the arguments mentioned before, it may be worthwhile to consider modified particle-physics
processes in the unitary sectors. An analysis of cosmic-ray data is likely to lead to an additional number
of tight constraints on Lorentz violation that complement the already existing dimension-five photon
sector bounds in the data tables \cite{KostRussell}.

\begin{acknowledgments}
The authors are grateful to CNPq, CAPES, and FAPEMA (Brazilian research agencies)
for financial support. In particular, M.M. Ferreira is obliged to
FAPEMA Universal 00880/15; FAPEMA PRONEX 01452/14; CNPq Produtividade 308933/2015-0. R.V.
Maluf acknowledges CNPq Produtividade 307556/2018-2, and M. Schreck is thankful to FAPEMA
Universal 01149/17; CNPq Universal 421566/2016-7; CNPq Produtividade 312201/2018-4.
In addition, we thank Z.~Li for pointing out the connection of
our modifications with a subset of those compiled in \cite{NMkost2} and for helpful
discussions. We are also indebted to the comments of the two anonymous referees that
led to an improved discussion of the motivation of our work and the results obtained.

\end{acknowledgments}

\begin{appendix}

\section{Comparison to propagator of MCFJ theory}
\label{sec:comparison-propagator-mcfj-theory}

It would be interesting to compare the propagator (\ref{eq:propagator-momentum-space}) with
that of MCFJ theory, which will be carried out in the current section. The MCFJ Lagrange density
is
\begin{equation}
\mathcal{L}_{\mathrm{MCFJ}}=-\frac{1}{4}F_{\alpha\beta}F^{\alpha\beta}-\frac{1}{4}%
\varepsilon^{\beta\alpha\rho\varphi}V_{\beta}A_{\alpha}F_{\rho\varphi}+\frac{1}{2\xi}(\partial_{\mu }A^{\mu})^{2}\,,
\end{equation}%
where $V_{\beta}$ is the CFJ vector background and the last term is included
to fix the gauge. Such a Lagrange density can be written as%
\begin{subequations}
\begin{align}
\mathcal{L}_{\mathrm{MCFJ}}&=\frac{1}{2}A^{\beta }\boxplus_{\beta\alpha}A^{\alpha }\,, \displaybreak[0]\\[2ex]
\boxplus_{\beta\alpha}&=\square\Theta_{\beta\alpha}-\frac{1}{\xi}\square\Omega_{\beta\alpha}+S_{\beta\alpha}\,,\quad S^{\beta\alpha}=\varepsilon^{\beta\alpha\varphi\rho }V_{\varphi}\partial_{\rho}\,,
\end{align}
\end{subequations}
with the projectors of Eq.~(\ref{Proj1}). Using an algebra similar to that
of Tab.~\ref{tab:closed-algebra}, one obtains the following propagator:%
\begin{align}
\Delta_{\alpha\nu}&=\frac{1}{\square\left[\square^{2}-(
V^{2}\square-\lambda^{2})\right]}\left[\square^{2}\Theta_{\alpha\nu}-\left\{\xi \left[\square^{2}-(V^{2}\square
-\lambda^{2})\right]+\lambda^{2}\right\}\Omega_{\alpha\nu}\right. \notag \\
&\hspace{3.8cm}\left.-\,\square(S_{\alpha\nu}+V_{\alpha}V_{\nu})+\lambda(V_{\alpha}\partial_{\nu}+V_{\nu}\partial_{\alpha})\right]\,,
\end{align}%
with $\lambda\equiv V^{\mu}\partial_{\mu}$. In momentum space, the latter is
\begin{subequations}
\begin{align}
\label{eq:propagador-MCFJ-theory}
\Delta_{\alpha \nu }(p)&=-\frac{\mathrm{i}}{p^{2}\left[p^{4}-(V\cdot
p)^{2}+V^{2}p^{2}\right]}\left[p^{4}\Theta_{\alpha\nu}\left(
p\right)-\left\{\xi\left[ p^{4}-\left( (V\cdot p)
^{2}-V^{2}p^{2}\right) \right] -(V\cdot p) ^{2}\right\} \frac{%
p_{\alpha}p_{\nu}}{p^{2}}\right.   \notag \\
&\phantom{{}={}}\hspace{4.2cm}\left.+\,p^{2}S_{\alpha\nu}(p)+p^{2}V_{\alpha}V_{\nu
}-(V\cdot p)(V_{\alpha}p_{\nu }+V_{\nu}p_{\alpha
})\right]\,, \displaybreak[0]\\
\Delta_{\alpha \nu }(p)&=-\frac{\mathrm{i}}{p^{2}}\left[1-\frac{(V\cdot
p)^{2}}{p^{4}}+\frac{V^{2}p^{2}}{p^{4}}\right]^{-1}\left\{\eta_{\alpha\nu}-\left[1-\frac{(V\cdot p)^{2}}{p^{4}}+\xi %
\left(1-\frac{(V\cdot p)^{2}}{p^{4}}+\frac{V^{2}p^{2}}{p^{4}}%
\right)\right]\frac{p_{\alpha}{}p_{\nu}}{p^{2}}\right. \notag \\
&\phantom{{}={}}\hspace{0.8cm}\left.+\,\frac{S_{\alpha\nu}(p)}{p^{2}}+\frac{V_{\alpha
}{}V_{\nu }}{p^{2}}-\frac{V\cdot p}{p^{4}}(V_{\alpha}p_{\nu}+V_{\nu}p_{\alpha})\right\}\,.
\end{align}
\end{subequations}
The replacement
\begin{equation}
\frac{V^{\mu}}{p^{2}}\mapsto 2D^{\mu},
\end{equation}%
implies
\begin{subequations}
	\begin{align}
	1-\frac{(V\cdot p)^{2}}{p^{4}}+\frac{V^{2}p^{2}}{p^{4}}&\mapsto 1-4(D\cdot p)^{2}+4D^{2}p^{2}\,, \\[2ex]
	\frac{S_{\alpha\nu}(p)}{p^{2}}&\mapsto 2L^{\beta\alpha}(p)\,,
	\end{align}
\end{subequations}%
which, inserted into Eq.~(\ref{eq:propagador-MCFJ-theory}), leads to the propagator~(\ref{eq:propagator-momentum-space})
of the first dimension-five model examined. Furthermore, this propagator, as it stands, corresponds to Eq.~(3.3) of the first
paper of Ref.~\cite{Adam2} for $V^{\mu}\mapsto mk_{\mu}$, $\xi\mapsto -\xi$ and $n^{\mu}\mapsto p^{\mu}$, with an
appropriate choice of the prefactor. Here, $m$ is the Chern-Simons mass and $n^{\mu}$ a fixed four-vector used in
their axial gauge fixing condition.

\section{Mapping to the nonminimal SME}
\label{sec:mapping-nonminimal-sme}

The modifications that we considered in this paper are as follows
\begin{subequations}
\begin{align}
\mathcal{L}_1&=\frac{1}{2}\epsilon^{\kappa\lambda\mu\nu}D_{\kappa}A_{\lambda}\square F_{\mu\nu}\,, \\[2ex]
\mathcal{L}_2&=\frac{1}{2}\epsilon^{\kappa\lambda\mu\nu}A_{\lambda}T_{\kappa}(T\cdot\partial)^{2}F_{\mu\nu}\,,
\end{align}
\end{subequations}
with the vector-valued background fields $D_{\mu}$ and $T_{\mu}$. We would like to map these Lagrangians onto those presented in~\cite{NMkost2}. As the related field operators are of mass dimension 5, the correct Lagrangian must be $\mathcal{L}_A^{(5)}$ of their Tab.~III with the observer tensor $k^{(5)\alpha\varrho\lambda\mu\nu}$. The pieces antisymmetric in $\varrho$, $\lambda$ and $\mu$, $\nu$ contribute only. By performing a partial integration and neglecting the surface terms, $\mathcal{L}_A^{(5)}$ can be brought into a form more suitable for us:
\begin{equation}
\mathcal{L}_A^{(5)}=-\frac{1}{4}k^{(5)\alpha\varrho\lambda\mu\nu}F_{\varrho\lambda}\partial_{\alpha}F_{\mu\nu}=-\frac{1}{2}k^{(5)\alpha\varrho\lambda\mu\nu}\partial_{\varrho}A_{\lambda}\partial_{\alpha}F_{\mu\nu}=\frac{1}{2}k^{(5)\alpha\varrho\lambda\mu\nu}A_{\lambda}\partial_{\varrho}\partial_{\alpha}F_{\mu\nu}\,.
\end{equation}
Comparing the latter to the previous Lagrange densities leads to the correspondences
\begin{subequations}
\label{eq:mapping-coefficients}
\begin{align}
k^{(5)\alpha\varrho\lambda\mu\nu}&=\eta^{\alpha\varrho}\varepsilon^{\kappa\lambda\mu\nu}D_{\kappa}\,, \\[2ex]
k^{(5)\alpha\varrho\lambda\mu\nu}&=T^{\alpha}T^{\varrho}\varepsilon^{\kappa\lambda\mu\nu}T_{\kappa}\,,
\end{align}
\end{subequations}
for $\mathcal{L}_1$ and $\mathcal{L}_2$, respectively. We see that both are symmetric in $\alpha$, $\varrho$ due to the symmetry of the two spacetime derivatives. The problem is that the tensors are not antisymmetric in $\rho$, $\lambda$. Hence, they should be antisymmetrized by hand. For example, for the first Lagrangian, we obtain
\begin{equation}
k^{(5)\alpha\varrho\lambda\mu\nu}=\frac{1}{2}\left[\eta^{\alpha\varrho}\varepsilon^{\kappa\lambda\mu\nu}D_{\kappa}-\eta^{\alpha\lambda}\varepsilon^{\kappa\varrho\mu\nu}D_{\kappa}\right]\,.
\end{equation}
Inserting the latter into $\mathcal{L}_A^{(5)}$ after performing the partial integration leads to
\begin{align}
\mathcal{L}_A^{(5)}&=\frac{1}{4}\left[\eta^{\alpha\varrho}\varepsilon^{\kappa\lambda\mu\nu}D_{\kappa}-\eta^{\alpha\lambda}\varepsilon^{\kappa\varrho\mu\nu}D_{\kappa}\right]A_{\lambda}\partial_{\varrho}\partial_{\alpha}F_{\mu\nu} \notag \\
&=\frac{1}{4}\left[\varepsilon^{\kappa\lambda\mu\nu}D_{\kappa}A_{\lambda}\square F_{\mu\nu}-\varepsilon^{\kappa\varrho\mu\nu}D_{\kappa}A_{\lambda}\partial_{\varrho}\partial^{\lambda}F_{\mu\nu}\right] \notag \\
&=\frac{1}{4}\left[\varepsilon^{\kappa\lambda\mu\nu}D_{\kappa}A_{\lambda}\square F_{\mu\nu}-2D_{\kappa}A_{\lambda}\partial^{\lambda}\partial_{\varrho}\widetilde{F}^{\kappa\varrho}\right]\,,
\end{align}
with the dual electromagnetic field strength tensor $\widetilde{F}^{\mu\nu}$. The second contribution vanishes due to the homogenous Maxwell equations $\partial_{\varrho}\widetilde{F}^{\varrho\kappa}=0$ that are still valid in the presence of Lorentz violation. Hence, it is not necessary to antisymmetrize the tensors in $\varrho$, $\lambda$ by hand, whereby the mappings of Eqs.~(\ref{eq:mapping-coefficients}) are valid as they stand. The same argument holds for the second modification.

\end{appendix}

\end{document}